\input ./style/arxiv-general.cfg
\documentclass[aoas,MSNbibl,nameyear,seceqn,dvips]{arximspdf}
\makeatletter
   \@ifpackageloaded{graphicx}{}{\usepackage{graphicx}}
\makeatother

\doi{10.1214/13-AOAS700} 
\volume{9}
\issue{2}
\pubyear{2015}
\firstpage{821}
\lastpage{848}
\docsubty{FLA}

\makeatletter
\newcommand{\eqref}[1]{(\ref{#1})}
\newcommand{\V}[1]{\mathbf{#1}}
\newcommand{\VV}[1]{\bolds{#1}}
\newcommand{\M}[1]{\mathbf{#1}}
\newcommand{\MM}[1]{\bolds{#1}}
\newcommand{\F}[1]{\mathrm{#1}}
\newcommand{\sgn}{\operatorname{sgn}}
\newcommand{\T}[1]{{#1}^{\mathsf{T}}}

\newtheorem{theorem}{Theorem}
\newtheorem{lemma}{Lemma}
\newtheorem{proposition}{Proposition}

\newproclaim{definition}{Definition}
\newproclaim{remark}{Remark}
\makeatother

\begin{document}
\begin{frontmatter}

\title{Estimating heterogeneous graphical models for discrete data
with an application to roll call~voting}
\runtitle{Estimating heterogeneous graphical models}

\begin{aug}
\author[A]{\fnms{Jian} \snm{Guo}\thanksref{U1}\ead[label=e1]{jguo@hsph.harvard.edu}},
\author[B]{\fnms{Jie} \snm{Cheng}\thanksref{U2}\ead[label=e5]{jieche@umich.edu}},
\author[B]{\fnms{Elizaveta} \snm{Levina}\thanksref{U2,T1}\ead[label=e2]{elevina@umich.edu}},
\author[B]{\fnms{George}~\snm{Michailidis}\corref{}\thanksref{U2,T2}\ead[label=e3]{gmichail@umich.edu}}
\and
\author[B]{\fnms{Ji} \snm{Zhu}\thanksref{U2,T3}\ead[label=e4]{jizhu@umich.edu}\ead[label=u1,url]{http://www.foo.com}}
\thankstext{T1}{Supported in part by NSF Grants DMS-01-106772 and DMS-11-59005.}
\thankstext{T2}{Supported in part by NIH Grant 1RC1CA145444-0110 and NSF Grant DMS-12-28164.}
\thankstext{T3}{Supported in part by NSF Grants DMS-07-05532 and DMS-07-48389.}

\runauthor{J. Guo et al.}

\affiliation{Harvard University\thanksmark{U1} and University of
Michigan\thanksmark{U2}}

\address[A]{J. Guo\\
Department of Biostatistics\\
Harvard University\\
655 Huntington Avenue\\
Boston, Massachusetts 02115\\
USA\\
\printead{e1}}

\address[B]{J. Cheng\\
E. Levina\\
G. Michailidis\\
J. Zhu\\
Department of Statistics\\
University of Michigan\\
1085 South University\\
Ann Arbor, Michigan 48109-1107\\
USA\\
\printead{e5}\\
\phantom{E-mail:\ }\printead*{e2}\\
\phantom{E-mail:\ }\printead*{e3}\\
\phantom{E-mail:\ }\printead*{e4}}
\end{aug}

\received{\smonth{5} \syear{2013}}
\revised{\smonth{10} \syear{2013}}

%
\begin{abstract}
We consider the problem of jointly estimating a collection
of graphical models for discrete data, corresponding to several
categories that share some common structure. An example for such a
setting is voting records of legislators
on different issues, such as defense, energy, and healthcare. We
develop a Markov graphical model to characterize the heterogeneous
dependence structures arising from such data. The model is fitted via a
joint estimation method that preserves the underlying common graph
structure, but also allows for differences between the networks. The
method employs a group penalty that targets the common zero interaction
effects across all the networks. We apply the method to describe the
internal networks of the U.S. Senate on several important issues. Our
analysis reveals individual structure for each issue, distinct from the
underlying well-known bipartisan structure common to all categories
which we are able to extract separately. We also establish consistency
of the proposed method both for parameter estimation and model
selection, and evaluate its numerical performance on a number of
simulated examples.
\end{abstract}

%
\end{frontmatter}

\section{Introduction}\label{sec_introduction}

The analysis of roll call data of legislative bodies has
attracted a lot of attention both in the political science and statistical
literature. For political scientists, such data allow to study broad
issues such as party cohesion as well as more specific ones such as
coalition formation; see, for example, the books by \citet
{EnelowHinich84,MatthewsStimson75,Morton99,PooleRosenthal97}.
A popular tool in political science is the ideal point model [\citet
{ClintonJackmanRivers04}] that posits a one-dimensional latent political
space along which legislators and bills they vote for are aligned.
A~legislator's position corresponds to an ideal point, where bills
coinciding with that position maximize his/her utility. These ideal
points reveal legislators' preferences and it is of interest to
infer them from roll call data. An extension of this model that
incorporates information about the text of the bills being voted upon
is discussed in \citet{Gerrish2011}, while the impact of absenteeism is
examined in \citet{Han2007}.

\begin{figure}

\includegraphics{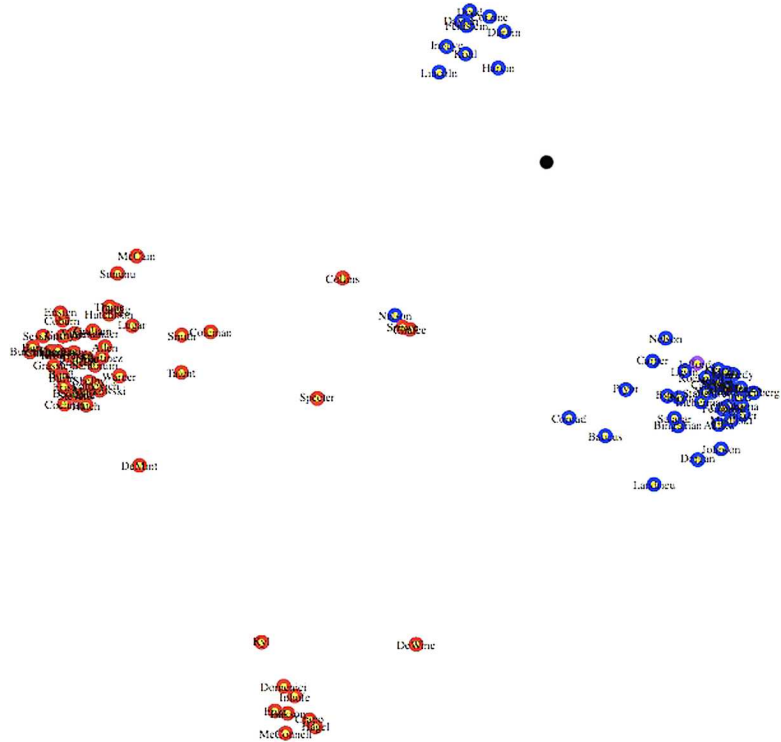}

\caption{Multidimensional scaling projection of
roll call data of the U.S. Senate for the period 2005--2006 (Republicans
shown in red and Democrats in blue).} \label{mds-results}
\end{figure}

A statistical challenge is how to best model and present the roll call
data in a way that makes interesting patterns apparent and facilitates
subsequent analyses. A number of techniques have been employed
including principal components analysis (PCA) [\citet{deLeeuw06}],
multidimensional scaling (MDS) [\citet{DiaconisGoelHolmes08}], Bayesian
spatial voting models [\citet{ClintonJackmanRivers04}], and
graphical models for binary data [\citet{BanerjeeGhaouidAspremont08}].

Dimension reduction techniques such as PCA and MDS aim at constructing
a ``map,'' with the
members of the legislative body positioned relative to their peers
according to their voting pattern.
A typical example of such a map of the
U.S. Senate members in the 109th Congress (2005--2006)
using multidimensional scaling for selected votes is shown in
Figure~\ref{mds-results}; for a detailed
description of the data see Section~\ref{sec_real_example}. A clear
separation between
members of the two parties is seen (Republicans to the left of the map
and Democrats to the right), together with some members exhibiting a
voting pattern deviating from their party, for example,
Nelson (Democrat of Nebraska), and Collins and Snow (Republicans of
Maine), while the independent Jeffords (shown in purple) votes like a
Democrat. More interestingly, the voting patterns within both parties
form distinct subclusters. While the nature of this division is
impossible to infer from an MDS or a PCA representation such as the one
shown in Figure~\ref{mds-results}, our subsequent analysis will show
that this difference is driven
by votes on defense/security and healthcare issues.

This finding suggests that treating all votes as homogeneous, that is,
assuming that they represent the same underlying relationship
between senators, may mask more subtle patterns which depend on the
issues being voted upon. Therefore, treating votes as heterogeneous is
more accurate and can provide further insight into the voting behavior
of different groups of senators on different issues. In this paper, we
focus on voting records on three types of bills: defense and national
security, environment and energy, and healthcare issues. Voting on the
latter category is typically more partisan than voting on defense and
national security and, thus, we expect to see different connections in
different categories.

The voting records of the U.S. Senate from the 109th Congress covering
the period 2005--2006 were obtained directly from the Senate's website
(\surl{www.senate.gov}). We chose the 109th Congress because its voting
patterns have been previously analyzed in the literature [see, e.g.,
\citet{BanerjeeGhaouidAspremont08}], but as we have discovered, the
version of the data previously analyzed was contaminated with voting
records from the 1990s (when the set of senators would have been
different). Thus, we collected the data ourselves, on all the 645 votes
that the Senate deliberated and voted on during that period, which
include bills, resolutions, motions, debates and roll call votes. To
study the potential heterogeneity in the voting patterns, we focused on
the three largest meaningful (i.e., excluding purely procedural votes)
categories of votes extracted from bills, resolutions and motions:
(1) defense and security issues; (2) environment and energy issues; (3)
health and medical care issues. The categories were extracted by a
combination of text analysis of bill names and manual labeling.
A~complete analysis of this data set will be presented in Section~\ref
{sec_real_example}.

Our goal in this paper is to develop a statistical model for studying
dependence patterns in such situations: there is some overall structure
present (party affiliation, which affects everything) and there are
also distinct categories with their own individual structures. Since we
are dealing with voting data, we use Markov network models to capture
the dependence structure of binary or categorical random variables.
Similar to Gaussian graphical models,
nodes in a Markov network correspond to (categorical) variables, while
edges represent dependence between nodes conditional on all other variables.
Graphical models are an exploratory data analysis tool used in a number
of application areas to explore the dependence structure between
variables, including bioinformatics [\citet{Airoldi07}], natural language
processing [\citet{JungParkChoiKim96}] and image analysis [\citet{Li01}].
In the case of Gaussian graphical models, which assumes the variables
are jointly normally distributed, the structure of the underlying graph
can be fully determined from the corresponding inverse covariance
(precision) matrix, the off-diagonal elements of which are proportional
to partial correlations between the variables. A number of methods have
been recently proposed in the literature to fit \emph{sparse} Gaussian
graphical models [see, e.g.,
\citet
{MeinshausenBuhlmann06,YuanLin07,BanerjeeGhaouidAspremont08,RothmanBickelLevinaZhu08,
RavikumarWainwrightRaskuttiYu08,PengWangZhouZhu09} and references
therein]. Sparse Markov networks for binary data (Ising models) have
been studied by \citet
{HoeflingTibshirani09,GuoLevinaMichailidisZhu09jointIsingtechreport,
RavikumarWainwrightLafferty10, Anandkumar12, Xue12}. These methods do
not allow for different categories within the data.

To allow for heterogeneity, we develop a framework for fitting
different Markov models for each category that are nevertheless \emph
{linked}, sharing nodes and some common edges across all categories,
while other edges are uniquely associated with a particular category.
This will allow us to borrow strength across categories instead of
fitting them completely separately. For the Gaussian case, this type of
joint graphical model was first studied by \citet
{GuoLevinaMichailidisZhu09CGM}, who proposed a joint likelihood based
estimation method that borrowed strength across categories. Several
other papers have proposed alternative algorithms for the Gaussian case
[\citet{Danaher,Yang12,Hara201323}]. We note that a context-specific
graphical model was proposed for count data in the form of contingency
tables by \citet{Hojsgaard04}, but contingency tables are not suitable
for high-dimensional data and the context-specific model is not sparse.

The advantage of using a Markov graphical model in this context is that
it quantifies the degree of conditional dependence between the senators based
on their voting record, and hence the obtained network, and is directly
interpretable. Techniques like multidimensional
scaling and principal components analysis represent relative
similarities between senators' voting records on the map and, hence,
the distance between any two senators can be interpreted as a
quantitative measure of similarity between their
voting records. However, unlike in a Markov network, these distances
are not interpretable in the context of a generative probability model.

The remainder of the paper is organized as follows. Section~\ref
{sec_methodology} introduces the Markov network and addresses
algorithmic issues, and Section~\ref{sec_simulation} briefly
illustrates the performance of the joint estimation method on simulated
data. A detailed analysis of the U.S. Senate's
voting record from the 109th Congress is presented in Section~\ref
{sec_real_example}. Some concluding remarks are drawn in Section~\ref
{sec_conclusion}, and the \hyperref[app]{Appendix} presents results on the asymptotic
properties of the method. The electronic supplementary material
contains a detailed investigation of missing data imputation methods
for the Senate vote data.

\section{Model and estimation algorithm}\label{sec_methodology}
In this section we present the Markov model for heterogeneous data,
focusing on the special case of binary variables (also known as the
Ising model).
The extension to general categorical variables is briefly discussed in
Section~\ref{sec_conclusion}. We start by discussing estimation of
\emph{separate} models for
each category and then develop a method for joint estimation.

The main technical challenge when estimating the likelihood of Markov
graphical models is its computational intractability
due to the normalizing constant. To overcome this difficulty, different
methods employing computationally tractable approximations to the
likelihood have been proposed in the literature;
these include methods based on surrogate likelihood [\citet
{BanerjeeGhaouidAspremont08,KolarXing08}] and pseudo-likelihood [\citet
{HoeflingTibshirani09,RavikumarWainwrightLafferty10,GuoLevinaMichailidisZhu10JOSE}].
\citet{HoeflingTibshirani09} also proposed an iterative algorithm that
successively approximates the original likelihood through a series of
pseudo-likelihoods, while \citet{RavikumarWainwrightLafferty10} and
\citet{GuoLevinaMichailidisZhu10JOSE} established asymptotic
consistency of their respective methods.

\subsection{Problem setup and separate estimation}\label{sec_review_SE}
We start from setting up notation and reviewing previous work on
estimating a single Ising model, which can be used to estimate the
graph for each category separately. Suppose that data have been
collected on $p$ \emph{binary} variables in $K$ categories, with $n_k$
observations in the $k$th category, $k = 1, \ldots, K$.
Let $\V{x}_{i}^{(k)} = (x_{i,1}^{(k)},\ldots, x_{i,p}^{(k)})$ denote
a $p$-dimensional row vector containing the data for the $i$th
observation in the $k$th category and
assume that it is drawn independently from an exponential family with
the probability mass function
%
\begin{equation}\qquad
\label{joint_density_Ising} \F{f}_k(X_1,\ldots, X_p) =
\frac{1}{\F{Z}(\MM{\Theta}^{(k)})} \exp \Biggl(\sum_{j=1}^p
\theta_{j,j}^{(k)} X_j + \sum
_{1 \le j < j'
\le p} \theta_{j,j'}^{(k)} X_j
X_{j'} \Biggr).
\end{equation}
The partition function $\F{Z}(\MM{\Theta}^{(k)})= \sum_{X_j \in\{
0,1\}, j} \exp(\theta_{j,j}^{(k)} X_j + \sum_{j < j'} \theta
_{j,j'}^{(k)} X_j X_{j'})$ ensures that the probabilities
in \eqref{joint_density_Ising} add up to one.
The parameters $\theta_{j,j}^{(k)}$, $1 \le j \le p$ correspond to the
main effect for variable $X_j$ in the $k$th category, and
$\theta_{j,j'}^{(k)}$ is the interaction effect between variables
$X_j$ and $X_{j'}$, $1 \le j < j' \le p$. The underlying network
associated with the $k$th category is determined by the symmetric
matrix $\MM{\Theta}^{(k)} = (\theta_{j,j'}^{(k)})_{p \times p}$.
Specifically, if $\theta_{j,j'}^{(k)}=0$, then $X_j$ and $X_{j'}$ are
conditionally independent in the $k$th category given all the remaining
variables
and, hence, their corresponding nodes are \emph{not} connected. For
each category, \eqref{joint_density_Ising} is referred to as the
Markov network in the machine learning literature and as the log-linear
model in the statistics literature, where $\theta_{j,j'}^{(k)}$ is
also interpreted as the conditional log odds ratio between $X_j$ and
$X_{j'}$ given the other variables. Although general Markov networks
allow higher order interactions (3-way, 4-way, etc.), \citet
{RavikumarWainwrightLafferty10} pointed out that in principle one can
consider only the pairwise
interaction effects without loss of generality, since higher order
interactions can be converted to pairwise ones by introducing
additional variables
[\citet{WainwrightJordan08}]. For the rest of this paper, we only
consider models with pairwise interactions of the original binary variables.

The simplest way to deal with heterogenous data is to estimate $K$
separate Markov models, one for each category. If one further assumes
sparsity for the $k$th category, the structure of the underlying graph
can be estimated by regularizing the log-likelihood using an $\ell_1$ penalty:
%
\begin{eqnarray}
\label{separate_method} &&\max_{\MM{\Theta}^{(k)}} \frac{1}{n_k} \sum
_{i=1}^{n_k} \Biggl\{\sum_{j=1}^p
\theta_{j,j}^{(k)} x_{i,j}^{(k)} + \sum
_{j < j'} \theta _{j,j'}^{(k)}
x_{i,j}^{(k)} x_{i,j'}^{(k)} \Biggr\}
\nonumber
\\[-8pt]
\\[-8pt]
\nonumber
&&\qquad{}- \log
\F{Z}\bigl(\M {\Theta}^{(k)}\bigr) - \lambda\sum
_{j < j'} \bigl|\theta_{j, j'}^{(k)}\bigr|.
\end{eqnarray}
The $\ell_1$ penalty shrinks some of the interaction effects $\theta
_{j,j'}^{(k)}$ to zero and $\lambda$ controls the degree of sparsity.
However, estimating
\eqref{separate_method} directly is computationally infeasible due to
the nature of the partition function. A standard approach in such a
situation is to replace the likelihood with a pseudo-likelihood [\citet
{besag86}], which has been shown to work well in a range of situations. Here,
we use a pseudo-likelihood estimation method for Ising models [\citet
{HoeflingTibshirani09,GuoLevinaMichailidisZhu10JOSE}], based on
%
\begin{eqnarray}
\label{pseudo_separate_method}&& \max_{\MM{\Theta}^{(k)}}  \frac{1}{n_k} \sum
_{i=1}^{n_k} \sum_{j=1}^p
\biggl[x_{i,j}^{(k)} \biggl(\theta_{j,j}^{(k)}
+ \sum_{j' \neq j} \theta_{j,j'}^{(k)}
x_{i,j'}^{(k)} \biggr)
\nonumber
\\
&&\hspace*{46pt}\qquad{} - \log \biggl\{1 + \exp \biggl(\theta_{j,j}^{(k)} + \sum
_{j' \neq j} \theta_{j,j'}^{(k)}
x_{i,j'}^{(k)} \biggr) \biggr\} \biggr]
\\
&& \qquad{}- \lambda\sum_{j < j'} \bigl|\theta_{j, j'}^{(k)}\bigr|,\nonumber
\end{eqnarray}
where $\MM{\Theta}^{(k)}$ is restricted to be symmetric. Criterion
\eqref{pseudo_separate_method} can be efficiently maximized using the
modified coordinate descent algorithm of \citet{HoeflingTibshirani09}.

\subsection{Joint estimation of heterogeneous networks}
The separate estimation methods reviewed in the previous section do not
take advantage of the shared nodes among the categories and potential
common structure. Our goal here is to explicitly include this into the
estimation procedure. We start by reparameterizing each~$\theta
_{j,j'}^{(k)}$ as
%
\begin{equation}
\label{reparameterize_omega} \theta_{j,j'}^{(k)} = \phi_{j,j'}
\gamma_{j,j'}^{(k)}, \qquad 1 \le j \neq j' \le p; 1 \le k
\le K.
\end{equation}
To avoid sign ambiguities between $\phi_{j,j'}$ and $\gamma
_{j,j'}^{(k)}$, we restrict $\phi_{j,j'} \ge0$, $1 \le j < j' \le p$.
To preserve the symmetry of $\MM{\Theta}^{(k)}$, we also require $\phi
_{j,j'} = \phi_{j',j}$ and $\gamma_{j,j'}^{(k)}=\gamma
_{j',j}^{(k)}$, for all $1 \le j < j' \le p$ and $1 \le k \le K$.
Moreover, for identifiability reasons, we restrict the diagonal
elements $\phi_{j,j}=1$ and $\gamma_{j,j}^{(k)}=\theta_{j,j}^{(k)}$.\vspace*{1pt}
Note that $\phi_{j,j'}$ is a common factor across all $K$ categories
that\vspace*{-1pt} controls the occurrence of common links shared across categories,
while $\gamma_{j,j'}^{(k)}$ is an individual factor specific to the
$k$th category.
The proposed joint estimation method maximizes the following penalized
criterion:
%
\begin{eqnarray}
\label{model_cgm_eta12} \max_{\{\MM{\Phi}^{(k)}, \MM{\Gamma}^{(k)}\}_{k=1}^K} && \sum_{k=1}^K
\frac{1}{n_k} \sum_{i=1}^{n_k} \sum
_{j=1}^p \biggl[x_{i,j}^{(k)}
\biggl(\theta_{j,j}^{(k)} + \sum_{j' \neq j}
\theta_{j,j'}^{(k)} x_{i,j'}^{(k)} \biggr)
\nonumber
\\
&&\hspace*{43pt} \qquad{}- \log \biggl\{1 + \exp \biggl(\theta_{j,j}^{(k)} + \sum
_{j' \neq j} \theta_{j,j'}^{(k)}
x_{i,j'}^{(k)} \biggr) \biggr\} \biggr]
\\
&&\qquad{} - \eta_1 \sum_{j < j'}
\phi_{j,j'} - \eta_2 \sum_{j < j'}
\sum_{k=1}^K \bigl|\gamma_{j,j'}^{(k)}\bigr|,\nonumber
\end{eqnarray}
where $\MM{\Phi}^{(k)}=(\phi_{j,j'})_{p \times p}$ and $\MM{\Gamma
}^{(k)} = (\gamma_{j,j'}^{(k)})_{p \times p}$. The tuning parameter
$\eta_1$ controls sparsity of the common structure across the $K$
networks. Specifically, if $\phi_{j,j'}$ is shrunk to zero, all
$\theta_{j,j'}^{(1)},\ldots, \theta_{j,j'}^{(K)}$ are also zero
and, hence, there is no link between nodes $j$ and $j'$ in any of the
$K$ graphs. Similarly,
$\eta_2$ is a tuning parameter controlling sparsity of links in
individual categories. Due to the nature of the $\ell_1$ penalty, some
of $\gamma_{j,j'}^{(k)}$'s will be shrunk to zero,\vspace*{1pt} resulting in a
collection of graphs with individual differences. Note that this
two-level penalty was originally proposed by \citet{ZhouZhu07} for
group variable selection in linear regression.

The criterion \eqref{model_cgm_eta12} achieves the stated goal of
estimating common structure and hence borrows strength across the $K$
data categories,
but requires the selection of two tuning parameters. However, there is
an equivalent criterion presented next that only involves a single
tuning parameter,
thus simplifying the estimation task
%
\begin{eqnarray}
\label{model_cgm_lambda} &&\max_{\{\MM{\Theta}^{(k)}\}_{k=1}^K} \sum_{k=1}^K
\frac{1}{n_k} \sum_{i=1}^{n_k} \sum
_{j=1}^p \biggl[x_{i,j}^{(k)}
\biggl(\theta_{j,j}^{(k)} + \sum_{j' \neq j}
\theta_{j,j'}^{(k)} x_{i,j'}^{(k)} \biggr)
\nonumber
\\
&&\hspace*{79pt}\qquad{} - \log \biggl\{1 + \exp \biggl(\theta_{j,j}^{(k)} + \sum
_{j' \neq j} \theta_{j,j'}^{(k)}
x_{i,j'}^{(k)} \biggr) \biggr\} \biggr]
\\
&&\qquad{} - \lambda\sum_{1 \le j < j' \le p} \sqrt{\sum
_{k=1}^K \bigl| \theta _{j,j'}^{(k)}\bigr|},\nonumber
\end{eqnarray}
where $\lambda=2\sqrt{\eta_1 \eta_2}$.
The optimization problems given by \eqref{model_cgm_eta12} and \eqref
{model_cgm_lambda} are equivalent in the sense that for each pair of
$(\eta_1,\eta_2)$ there is a $\lambda$ that gives the same solution
and vice versa. Their equivalence
can be formalized as follows (here $\M{A} \cdot\M{B}$ denotes the
Schur--Hadamard element-wise product of two matrices):

\begin{proposition}\label{lemma_etatolambda}
Let $\{\widehat{\MM{\Theta}}^{(k)}\}_{k=1}^K$ be a local maximizer of
\eqref{model_cgm_lambda}. Then there exists a local maximizer of
\eqref{model_cgm_eta12}, $(\widehat{\MM{\Phi}}, \{\widehat{\M
{\Gamma}}^{(k)}\}_{k=1}^K)$, such that $\widehat{\MM{\Theta}}^{(k)}
= \widehat{\MM{\Phi}} \cdot\widehat{\MM{\Gamma}}^{(k)}$, for all
$1 \le k \le K$. On the other hand, if $(\widehat{\MM{\Phi}}, \{
\widehat{\MM{\Gamma}}^{(k)}\}_{k=1}^K)$ is a local maximizer of
\eqref{model_cgm_eta12}, then there also exists a local maximizer of
\eqref{model_cgm_lambda}, $\{\widehat{\MM{\Theta}}^{(k)}\}_{k=1}^K$,
such that $\widehat{\MM{\Theta}}^{(k)} = \widehat{\MM{\Phi}} \cdot
\widehat{\MM{\Gamma}}^{(k)}$, for all $1 \le k \le K$.
\end{proposition}

The proof of this proposition is similar to the proofs of Lemma~1 and
Theorem~1 in \citet{ZhouZhu07} and is omitted here. Note that even
though choosing a single tuning parameter $\lambda$ corresponds to a
particular path in the $(\eta_1, \eta_2)$ space, this restriction
affects only the individual estimates $\phi_{j,j'}$ and $\gamma
_{j,j'}$, but not their product $\theta_{j,j'}$.

\subsection{Algorithm and model selection}\label{sec_algorithm_models}
Criterion \eqref{model_cgm_lambda} leads to an efficient estimation
algorithm based on the local linear approximation. Specifically, 
letting $(\theta_{j,j'}^{(k)})^{[t]}$ denote the estimates from the
$t$th iteration, we approximate $\sqrt{\sum_{k=1}^K | \theta
_{j,j'}^{(k)}|} \approx\sum_{k=1}^K | \theta_{j,j'}^{(k)}| / \sqrt
{\sum_{k=1}^K |(\theta_{j,j'}^{(k)})^{[t]}|}$, when $\theta
_{j,j'}^{(k)} \approx(\theta_{j,j'}^{(k)})^{[t]}$.
Thus, at the $(t+1)$th iteration, problem \eqref{model_cgm_lambda} is
decomposed into $K$ individual
optimization problems:
%
\begin{eqnarray}
\label{model_LLA}&& \max_{\MM{\Theta}^{(k)}}  \frac{1}{n_k} \sum
_{i=1}^{n_k} \sum_{j=1}^p
\biggl[x_{i,j}^{(k)} \biggl(\theta_{j,j}^{(k)}
+ \sum_{j' \neq j} \theta_{j,j'}^{(k)}
x_{i,j'}^{(k)} \biggr)
\nonumber
\\
&&\hspace*{48pt}\qquad{} - \log \biggl\{1 + \exp \biggl(\theta_{j,j}^{(k)} + \sum
_{j' \neq j} \theta_{j,j'}^{(k)}
x_{i,j'}^{(k)} \biggr) \biggr\} \biggr]
\\
&& \qquad{}- \lambda\sum_{1 \le j < j' \le p} \Biggl(\sum
_{k=1}^K \bigl| \bigl(\theta _{j,j'}^{(k)}
\bigr)^{[t]}\bigr| \Biggr)^{-1/2} \bigl|\theta_{j,j'}^{(k)}\bigr|\nonumber.
\end{eqnarray}
Note that criterion
\eqref{model_LLA} is a variant of criterion \eqref
{pseudo_separate_method} with a weighted $\ell_1$ penalty and hence
can be solved by the algorithm of \citet{HoeflingTibshirani09}. For
numerical stability, we
threshold $\sqrt{\sum_{k=1}^K |(\theta_{j,j'}^{(k)})^{[t]}|}$ at
$10^{-10}$. The algorithm is summarized as follows:
\begin{longlist}[\textit{Step} 1.]
\item[\textit{Step} 1.] Initialize $\widehat{\theta}_{j,j'}^{(k)}$'s ($1 \le
j,j' \le p;1 \le k \le K$) using the estimates from the separate
estimation method;
\item[\textit{Step} 2.] For each $1 \le k \le K$, update $\widehat{\theta
}_{j,j'}^{(k)}$'s by solving \eqref{model_LLA} using the
pseudo-likelihood algorithm \citet
{HoeflingTibshirani09,GuoLevinaMichailidisZhu10JOSE}.
\item[\textit{Step} 3.] Repeat step 2 until convergence.
\end{longlist}

The tuning parameter $\lambda$ in \eqref{model_cgm_lambda} controls
the sparsity of the resulting estimator and can be selected using
cross-validation. Specifically, for each $1 \le k \le K$, we randomly
split the data in the $k$th category into $D$ subsets of similar sizes
and denote the index set of the observations in the $d$th subset as
$\mathcal{T}_{d}^{(k)}$, $1 \le d \le D$. Then $\lambda$ is selected
by maximizing
%
\begin{eqnarray}
\label{cv_cgm} &&\frac{1}{D} \sum_{d=1}^D
\sum_{k=1}^K \frac{1}{|\mathcal
{T}_{d}^{(k)}|} \sum
_{i \in\mathcal{T}_{d}^{(k)}} \sum_{j=1}^p
 x_{i,j}^{(k)} \biggl\{\bigl(\widehat{\theta}_{j,j}^{(k)}
\bigr)^{[-d]}(\lambda) + \sum_{j' \neq j} \bigl(
\widehat{\theta}_{j,j'}^{(k)}\bigr)^{[-d]}(\lambda)
x_{i,j'}^{(k)} \biggr\}
\nonumber
\\[-8pt]
\\[-8pt]
\nonumber
&&\qquad{}- \log \biggl[1 + \exp \biggl\{\bigl(\widehat{\theta }_{j,j}^{(k)}
\bigr)^{[-d]}(\lambda) + \sum_{j' \neq j} \bigl(
\widehat{\theta }_{j,j'}^{(k)}\bigr)^{[-d]}(\lambda)
x_{i,j'}^{(k)} \biggr\} \biggr],
\end{eqnarray}
where $|\mathcal{T}_{d}^{(k)}|$ is the cardinality of $\mathcal
{T}_{d}^{(k)}$ and $(\widehat{\theta}_{j,j'}^{(k)})^{[-d]}(\lambda)$
is the joint estimate of $\theta_{j,j'}^{(k)}$ based on all
observations except those in $\mathcal{T}_{d}^{(1)}\cup\cdots
\cup\mathcal{T}_{d}^{(K)}$, as well as the tuning parameter
$\lambda$.

\section{Simulation study}\label{sec_simulation}
\begin{figure}

\includegraphics{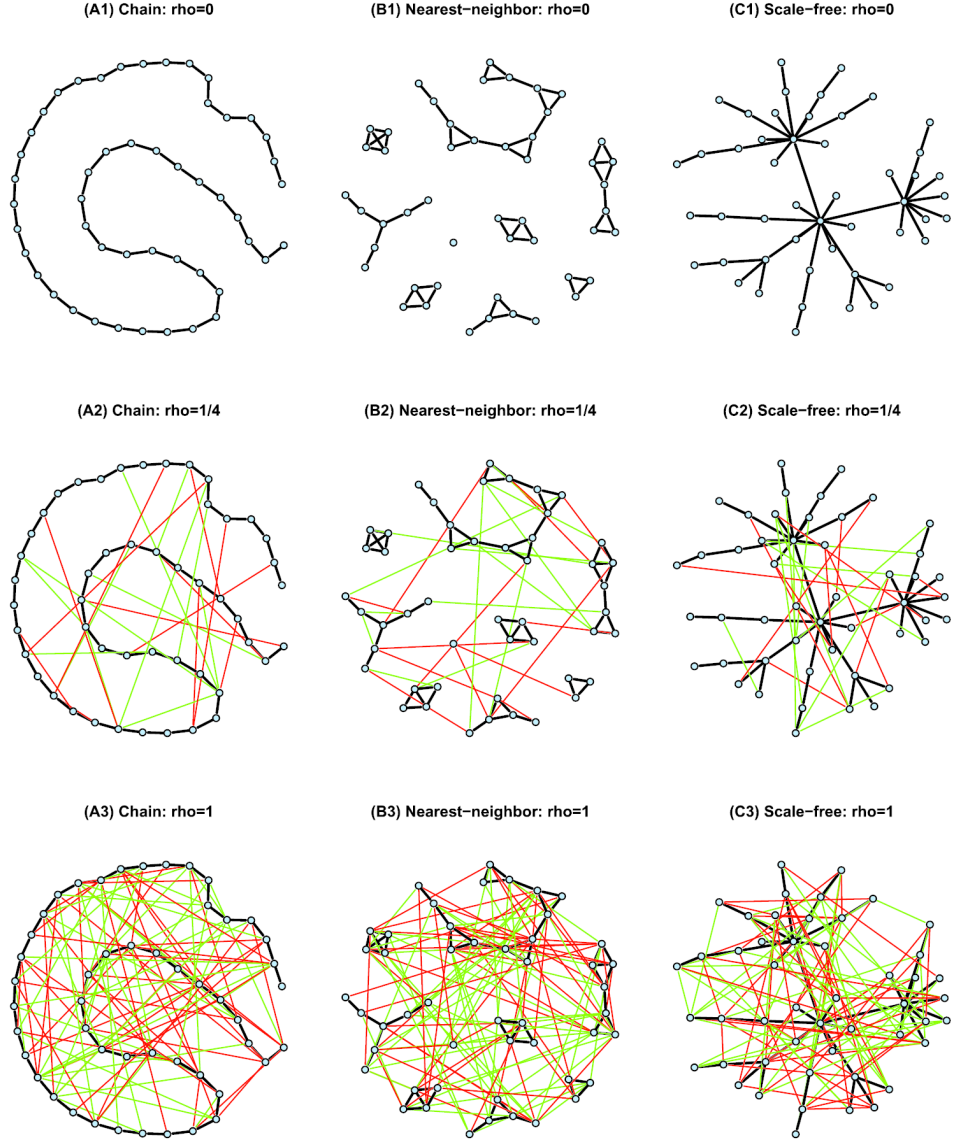}

\caption{The networks used in three simulated examples. The black
lines represent the common structure, whereas the red, blue and green
lines represent the individual links in the three categories. $\rho$
is the ratio of the number of
individual links to the number of common links.}\label{fig_illustrate}\vspace*{-5pt}
\end{figure}

Before turning our attention to examining the U.S. Senate voting patterns,
we evaluate the performance of the joint estimation method on three
synthetic examples, each with $p=100$ variables and $K=3$  categories.
The network structure in each example is composed of two parts:\vadjust{\goodbreak} the
common structure across all categories and the individual structure
specific to a category. The common structures in these examples are a
chain graph, a nearest neighbor graph and a scale-free graph. These
graphs are generated as follows:
\begin{longlist}[Example~1:]
\item[Example~1:] \textit{Chain graph}.
A chain graph is generated by connecting nodes 1 to~$p$ in increasing
order, as shown in Figure~\ref{fig_illustrate}(A1).
\item[Example~2:] \textit{Nearest neighbor graph.}
The data generating mechanism of the nearest neighbor graph is adapted
from \citet{LiGui06}. Specifically, we generate $p$ points randomly on
a unit square, calculate all $p(p-1)/2$ pairwise distances, and find
three nearest neighbors of each point in terms of these distances. The
nearest neighbor network is obtained by linking any two points that are
nearest neighbors of each other. Figure~\ref{fig_illustrate}(B1)
illustrates a nearest-neighbor graph.
\item[Example~3:] \textit{Scale-free graph.}
A scale-free graph has a power-law degree distribution and can be
simulated by the Barabasi--Albert algorithm\break [\citet{BarabasiAlbert99}].
A realization of a scale-free network is depicted in Figure~\ref
{fig_illustrate}(C1).
\end{longlist}
%
%

In each example, the network for the $k$th category ($k=1,\ldots,K$)
is created by randomly adding links to the common structure. The
individual links in different categories are disjoint and have the same
degree of sparsity, measured by $\rho$, the ratio of the number of
individual links to the number of common links. In particular, $\rho
=0$ corresponds to identical networks for all three categories. In the
simulation study, we consider $\rho=0$, $1/4$ and 1, gradually
increasing the proportion of individual links (Figure~\ref
{fig_illustrate}). Given the graphs, the symmetric parameter matrix $\M
{\Theta}^{(k)}$ is generated as follows. Each $\theta
_{j,j'}^{(k)}=\theta_{j',j}^{(k)}$ corresponding to an edge between
nodes $j$ and $j'$ is uniformly drawn from $[-1, -0.5]\cup[0.5,1]$,
whereas all other elements are set to zero. Then we generate the data
using Gibbs sampling. Specifically, suppose the $i$th iteration sample
has been drawn and is denoted as $(x_1^{(k)})^{[t]},\ldots,
(x_p^{(k)})^{[t]}$; then, in the $(t+1)$th iteration, we draw
$(x_j^{(k)})^{[t+1]}$, $1 \le j \le p$, from the Bernoulli distribution:
%
\begin{equation}
\bigl(x_j^{(k)}\bigr)^{[t+1]} \sim\operatorname{Bernoulli}
\biggl(\frac{\exp(\theta
_{j,j}^{(k)} + \sum_{j' \neq j} \theta_{j,j'}^{(k)}
(x_{j'}^{(k)})^{[t]})}{1 + \exp(\theta_{j,j}^{(k)} + \sum_{j' \neq
j} \theta_{j,j'}^{(k)} (x_{j'}^{(k)})^{[t]})} \biggr).
\end{equation}
To ensure that the simulated observations are close to i.i.d. samples
from the target distribution, the first 1,000,000 rounds are discarded
(burn-in) and the data are collected every 100 iterations from the
sampler. In the simulation study, we consider a balanced scenario and
an unbalanced scenario. The former consists of $n_k=300$ observations
in each category, whereas the latter has three unbalanced categories
with sample sizes $n_1=200$, $n_2=300$ and $n_3=400$.

\begin{figure}[t]

\includegraphics{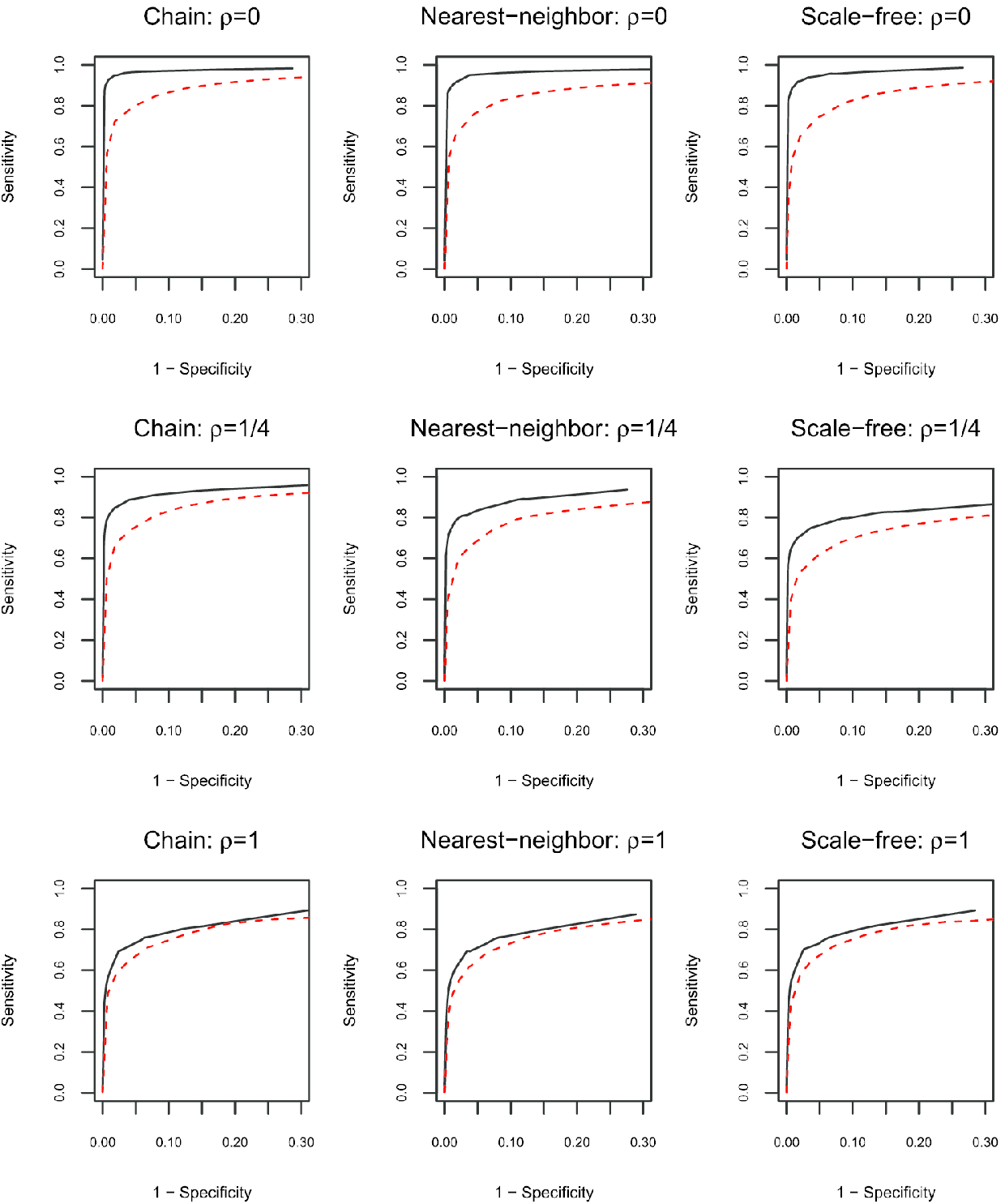}

\caption{Results for the balanced scenario ($n_1=n_2=n_3=300$) and
dimension $p = 100$. Black solid curve: joint estimation; red dashed
curve: separate estimation. The ROC curves are averaged over 10
replications. $\rho$ is the ratio between
the number of individual links and the number of common links.}\label
{fig_sim_subplot_K3}
\end{figure}

\begin{figure}[b]

\includegraphics{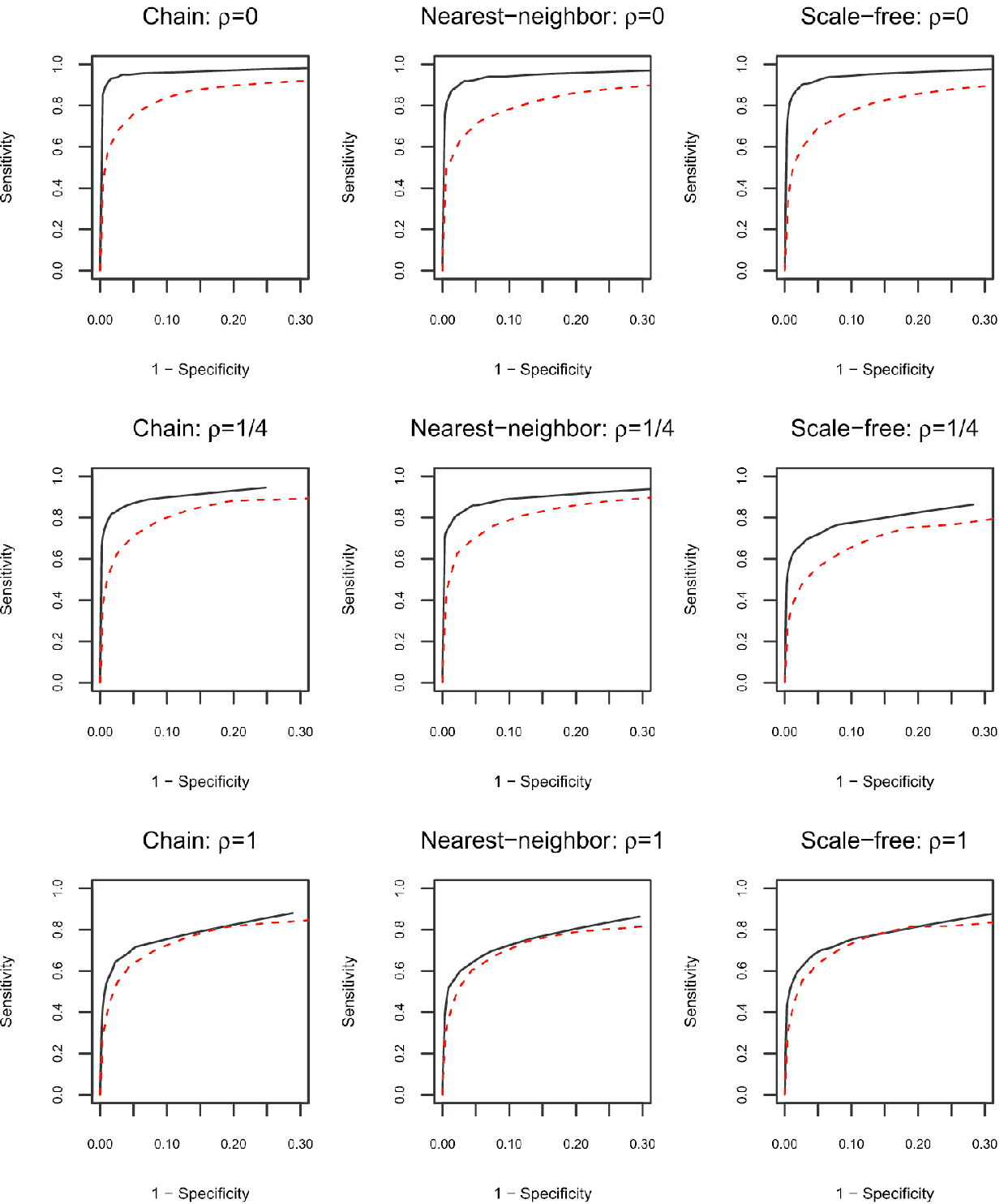}

\caption{Results for the unbalanced scenario ($n_1= 200$, $n_2=300$,
$n_3=400$) and dimension
$p = 100$. Black solid curve: joint estimation; red dashed curve:
separate estimation.The ROC curves are averaged over 10
replications. $\rho$ is the ratio between the number of individual
links and the number of common links.}\vspace*{-3pt}\label{fig_sim_subplot_unbalance}
\end{figure}

We compared the structure estimation results of the joint estimation
method and the separate estimation method using ROC curves, which
dynamically characterize the sensitivity (proportion of correctly
identified links) and the specificity (proportion of correctly excluded
links) by varying the tuning parameter $\lambda$. Figure~\ref
{fig_sim_subplot_K3} shows the ROC curves averaged over 10 replications
from the three examples in the balanced scenario, where the joint
estimation method
dominates separate estimation when the proportion of individual links
is low. As $\rho$ increases, the structures become more different, and
the joint and separate methods move closer together. This is expected,
since the joint estimation method is designed to take advantage of
common structure. The results in the unbalanced scenario exhibit a
similar pattern (Figure~\ref{fig_sim_subplot_unbalance}).\looseness=-1

\section{Analysis of the U.S. Senate voting records}
\label{sec_real_example}
We applied the proposed joint estimation method to the voting records
of the U.S. Senate from the 109th Congress covering the period
2005--2006. The $p=100$ variables correspond to the senators.
The Senate held 645 votes in that period, from which we extracted
$n=222$ votes in the three largest categories, namely, defense and
security (141), environment and energy (34), and healthcare (47).
The votes are recorded as ``yes'' (encoded as ``1'') and ``no''
(encoded as ``0''). The assumption of our model is that bills within a
category are an i.i.d. sample from the same underlying Ising model.
In reality, the voting process may be more complex, with possible
temporal factors and further dependencies among bills, possibly
reflecting backroom deals. Neverthless, this is an improvement on
previous analyses of such data, which treated all bills in all
categories as i.i.d. [\citet{BanerjeeGhaouidAspremont08}], and is a
reasonable trade-off for an exploratory data analysis tool.

There were missing observations, as not all senators vote on all bills.
The number of bills containing at least one missing vote was
98 out of 141 for defense and security, missing a total of 2.26\% of
all votes; 24 out of 34 for environment and energy, missing a total of
3.23\% of votes; and 20 out of 47 for healthcare, missing 2.38\% of all
votes. While the number of bills that are missing at least one
Senator's vote is relatively high, the overall proportion of missing
observations is quite low and, thus, we do not expect it to create a
major problem in the analysis. Nevertheless, we have investigated
multiple strategies for imputing the missing data in the electronic
supplement; specifically, we considered replacing the missing vote by
the party's majority, by the majority vote of the five most similar
Senators and, to test robustness to the imputation method, also by the
opposite party's majority and at random. We found that the main
conclusions of the analysis are not very sensitive to missing data
imputation methods.
In the subsequent analysis, we replace a missing vote for a Senator by
his/her party's majority vote on the bill; for the Independent Senator
Jeffords, we take the Democratic majority vote.
After the imputation, the bills with a ``yes/no'' proportion greater
than 90\% or less than 10\% were excluded from the analysis, as these
typically correspond to procedural votes. This left 97, 29 and 40 bills
in the three categories, respectively. Given that two of the
sample sizes are fairly small (29 and 40), we added an $\ell_2$
penalty with a small tuning parameter $\lambda_2=0.01$.
This approach, known as the elastic net, has been shown to help avoid
extremely sparse networks in such situations [\citet{ZouHastie05}].

The main tuning parameter for our method was selected through
cross-validation. Following \citet{LiGui06}, we used a bootstrap
procedure for
final edge selection, estimating the network for 100 bootstrap samples
of the same size, and only retained
edges that appeared more that $\alpha$ percent of the time. This
procedure is similar to stability selection [\citet{MeinshausenBuhlmann10}].

\begin{figure}[b]

\includegraphics{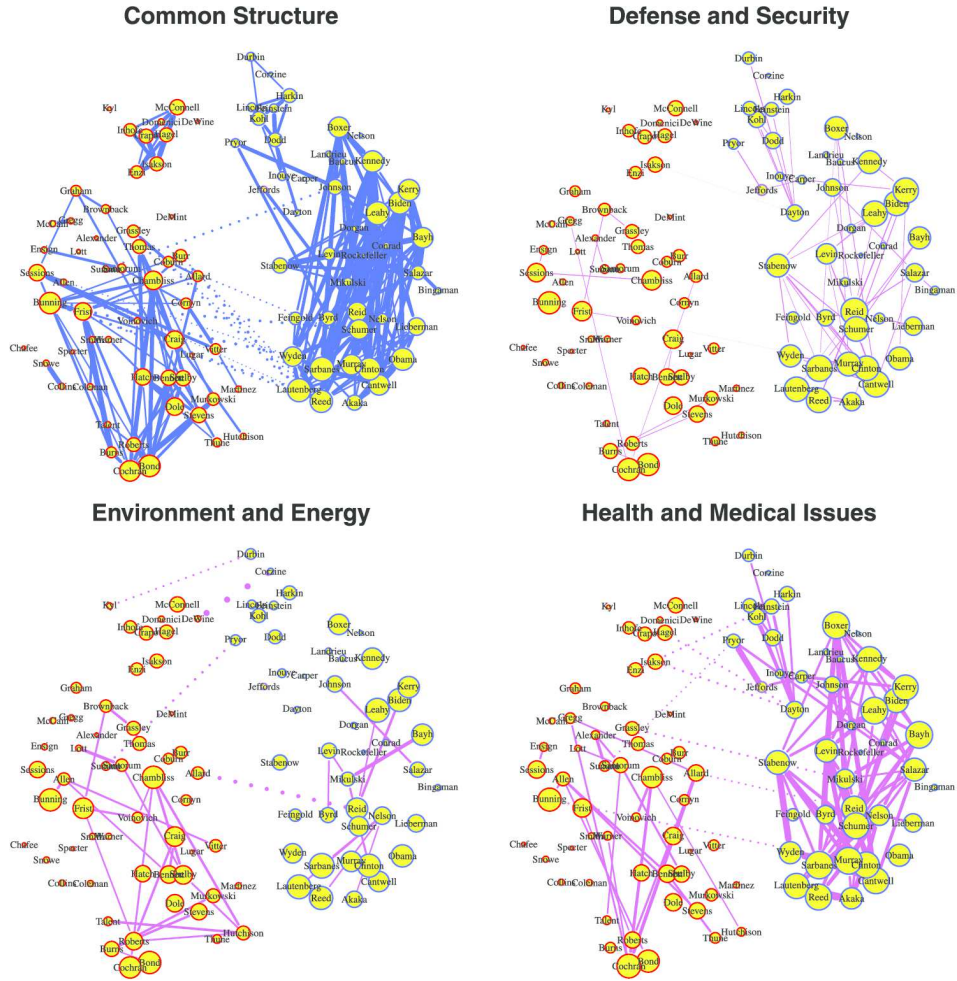}

\caption{The estimated graphical models for the three categories in
the Senate voting data with an inclusion cutoff value of
0.4 and tuning parameter value of 0.5. Edges common to all three
categories are shown under the heading
``common structure''; all other edges are shown on category-specific
graphs. The nodes represent the 100 senators,
with red, blue and purple node colors corresponding to Republican,
Democrat or Independent (Senator Jeffords),
respectively. A solid line corresponds to a positive interaction effect
and a dashed line to a negative interaction effect.
The width of a link is proportional to the magnitude of the
corresponding overall interaction effect.}
\label{fig_lambda05_cutoff04}
\end{figure}

The network representation, depicting both the common and the
individual structures with a cutoff value for inclusion
$\alpha=0.4$ and a value of $\lambda=0.05$, is depicted in
Figure~\ref{fig_lambda05_cutoff04}.
Note that unlike techniques such as principal components analysis and
multidimensional scaling that directly embed the senators in a
two-dimensional map, the proposed method estimates the edges and
constructs the adjacency matrix of the graph of Senators; subsequently,
we employed a graph drawing program to visualize this graph.
The common network structure estimated by the joint estimation method
is shown in the top left panel of Figure~\ref{fig_lambda05_cutoff04}.
For the individual
categories, we only plot the edges associated with the category that is
not part of the common network, to enhance the readability of the
graphs. As expected, members of the two political parties are clearly
separated. For both tuning parameter values, there are strong positive
associations between senators of the same party and selected strong
negative associations between senators of opposite parties. Obviously,
at the higher tuning parameter value the common dependence structure
becomes sparser. Of particular interest is the finding that at both
tuning values there are many more associations between Democratic
senators than Republican ones and
this pattern holds for both the common and individual structures. One
possible explanation may be that during that period the Democrats were
in the minority and thus voting more frequently as a block. Further,
the Independent
Senator Jeffords is associated with the Democrats, while the moderate
Republicans Collins, Snowe, Chafee and Specter (who switched to the
Democratic party in early 2009) are not strongly associated with their
Republican colleagues,
thus confirming results of previous analyses by
\citet{ClintonJackmanRivers04} and \citet{deLeeuw06} (albeit based on
data from the 105th Congress). The conservative Democrat Nelson
(Nebraska) is also not closely associated with his party, as well as
the very conservative Republican
de Mint (South Carolina). Also, the analysis suggests that Senator
Lieberman had a solid Democratic voting record before becoming an
Independent in 2008.

Other interesting patterns emerging from
the analysis are that the more moderate members of the two parties are
located closer to the center of their respective ``clouds'' (e.g.,
Warner, Frist,
Voinovich and Smith on the Republican side, and Levin, Reid, Mikulski
and Rockefeller on the Democratic side), the cluster of
economic conservatives on the Republican side (McConnell, Domenici,
Crapo, Inhofe),
the close ties of the liberal Democrats Kennedy, Boxer and Nelson
(Florida), the
close voting records of senators from the same state (Schumer and
Clinton from New York,
Murkowski and Stevens from Alaska, Snowe and Collins from Maine,
Cantwell and Murray from Washington). There is also a strong dependence
between Durbin, Corzine, Lincoln, Harkin and
Dodd on the Democratic side.

Examining the individual networks for the three categories shown in
Figure~\ref{fig_lambda05_cutoff04},
we note that additional positive associations among Democrats emerge,
primarily for defense and healthcare categories, thus
indicating a stronger ideological cohesion on these issues. Further, a
number of stable negative associations emerge in the environment and
healthcare categories, indicating a stronger ideological divide between
senators.

On defense, some additional strong ties emerge between more liberal
leaning Democrats (Stabenow, Biden, Leahy, Kerry, Boxer),
while a strong cluster on environmental issues arises between
Republican senators from energy producing states
(Murkowski and Stevens
from Alaska, Thune from South Dakota, Hutchison from Texas, but also
Bond from Missouri, Chambliss from Georgia, Craig from
Idaho and Roberts from Kansas with their unwavering support for
offshore drilling).
On health and medical issues, a number of additional strong positive
associations emerge among Democratic senators, possibly reflecting the
fact that the 109th Congress dealt with issues ranging
from veterans affairs, to medical malpractice to food safety and
especially on health savings accounts legislation to reduce
medical insurance costs.

Different imputation strategies for missing data were also examined and
the analysis results are given in Figures 1--3 in the Supplement
for the same values of the cutoff $\alpha$ and tuning parameter
$\lambda$. It can be
seen that similar patterns emerge, although alternative methods of
imputation may lead to the emergence of a few more
associations. Nevertheless, the main findings seem to be robust to the
examined choices of the imputation mechanism, although at very high
levels of absenteeism this may not hold [\citet{Han2007}].

\begin{figure}

\includegraphics{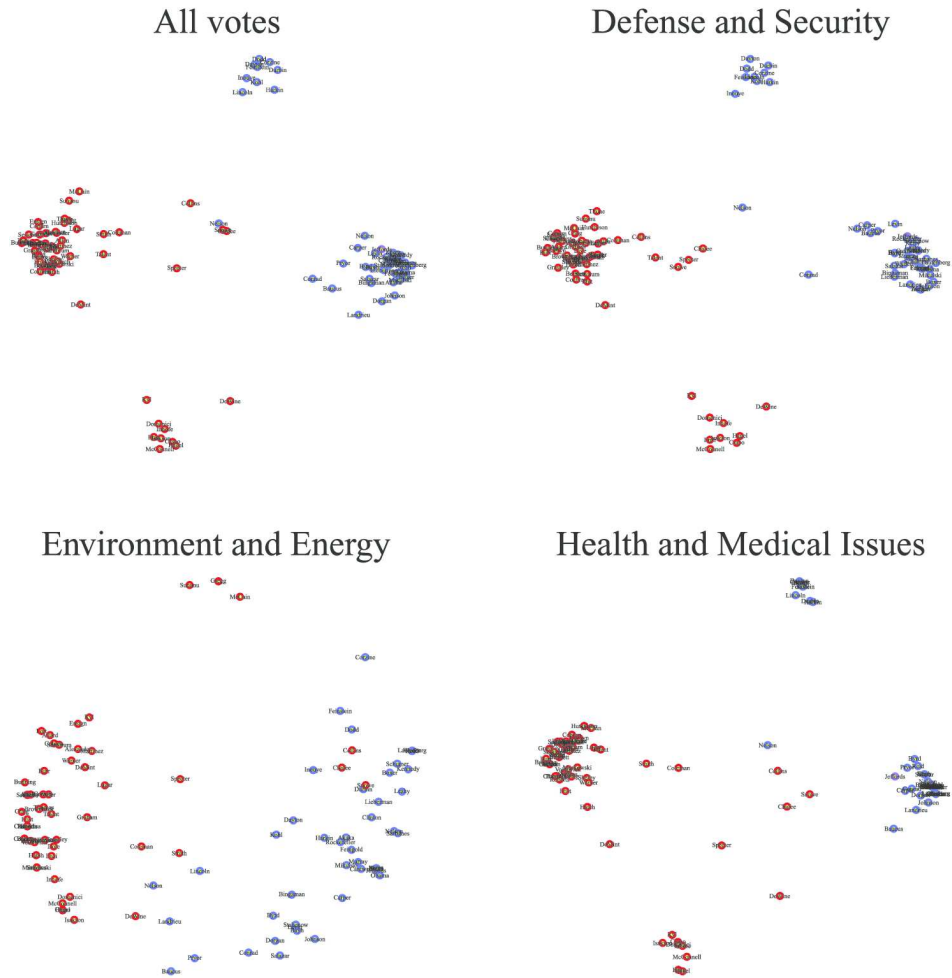}

\caption{Multidimensional scaling analysis for all the votes together,
and the three individual
categories. The nodes represent the 100 senators, with red, blue and
purple node colors corresponding to Republican,
Democrat or Independent (Senator Jeffords), respectively.}\label{mds-analyses}
\end{figure}

For comparison purposes, separate multidimensional scaling analyses are
shown in Figure~\ref{mds-analyses} for all the votes together and for
the three categories separately. MDS (or PCA or factor analysis) is one
of the commonly taken approaches in social sciences when graphical
modeling is not considered. Figure~\ref{mds-analyses} suggests that
the overall vote clustering in the two parties is driven to a large
extent by the corresponding clustering in the defense and health categories.
On the other hand, voting on environmental issues creates a clear
separation between the two parties, although the moderate Republicans
Chafee, Collins and Snowe are shown to have a voting record similar to
the Democrats, while the Democrats
Nelson (Nebraska) and Landrieu are closer to the Republicans. At a high
level, MDS-based findings are similar to ours, which is a satisfactory
result, but they do not provide explicit clusters or edges, nor do they
provide a way to quantify the amount of dependence between individual
pairs (visualized via edge thickness in Figure~\ref{fig_lambda05_cutoff04}).

\begin{figure}

\includegraphics{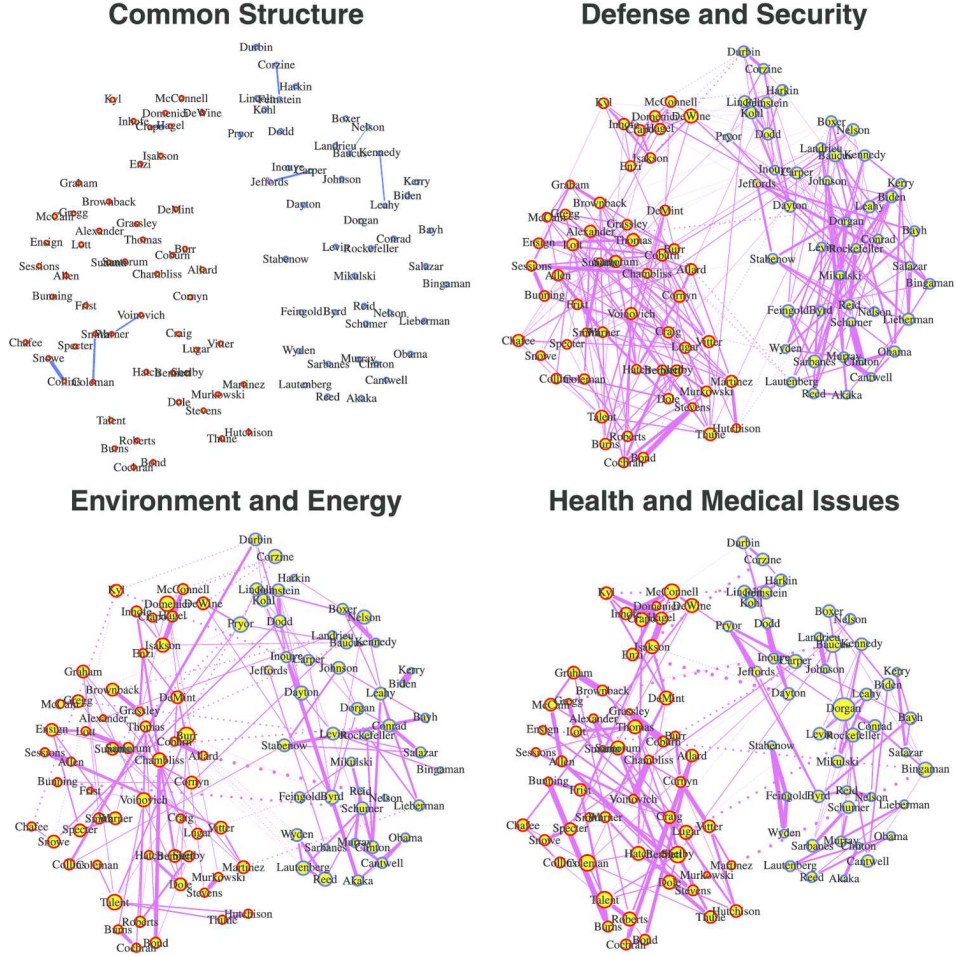}

\caption{The estimated graphical models for the three categories in
the Senate voting data fitted via separate estimation.
Edges common to all three categories are shown under the heading
``common structure''; all other edges
are shown on category-specific graphs. The cutoff value is 1 (only
edges appearing in all bootstrap
replications are included). The nodes represent the 100 senators, with
red, blue and purple node colors
corresponding to Republican, Democrat or Independent (Senator Jeffords),
respectively. A solid line corresponds to a positive interaction effect
and a dashed line to a negative
interaction effect. The width of a link is proportional to the
magnitude of the corresponding overall interaction effect.}\label
{separate-data-analysis}
\end{figure}

Another relevant comparison is to fitting a separate graphical model to
each of the three categories, as could have been done with any of the
previously developed methods for fitting the Ising model. The results
are shown in Figure~\ref{separate-data-analysis}, in the same format
as in Figure~\ref{fig_lambda05_cutoff04}, with edges common to all
three categories shown under ``common structure,'' and all other edges
under their own category. We followed the same tuning procedure as we
did for joint estimation, bootstrapping the data 100 times for
stability selection and selecting the value of the tuning parameter on
a validation data set. Even with the cutoff set at 1 (we included only
the edges appearing in all the bootstrap replications), the graphs are
dense and difficult to interpret. Similar to MDS, they capture party
cohesion through strong
positive associations between members of the same party for all three
categories and some negative associations between members of opposite
parties. However, different voting patterns between categories are not
clear, although the results suggest a more cohesive voting record for
both parties for the defense category. Note that since this is
exploratory data analysis, it is hard to verify which set of results is
``better.'' Nevertheless, those obtained from the joint estimation
method are more nuanced and interpretable and therefore provide
better insights into voting strategies of members of Congress.

\section{Concluding remarks}\label{sec_conclusion}
We have proposed a joint estimation method for the analysis of
heterogenous Markov networks motivated by the need to jointly estimate
heterogeneous networks, such as those of the Senate voting patterns.
The method improves estimation of the networks' common structure by borrowing
strength across categories, and allows for individual differences.
Asymptotic properties of the method have been established. In
particular, we show that the convergence rate is similar to the rate
for Gaussian graphical models in a similar context [\citet
{GuoLevinaMichailidisZhu10JOSE}]. The proposed method can be extended
to deal with general categorical data with more than two levels using
the strategy described in \citet{RavikumarWainwrightLafferty10} and
\citet{GuoLevinaMichailidisZhu10JOSE}.
The most interesting feature emerging from the analysis of the Senate
voting records is the existence of more stable associations for the
Democrats, both in terms of the common structure
and in the healthcare and defense categories.

There are other techniques suitable for analyzing roll call data.
Dimension reduction techniques
create maps, where the relative positioning of the senators allows one
to infer similarity in their voting patterns. They provide
a useful visual tool to capture broad patterns and relationships. On
the other hand, a Markov network model aims directly at estimating
the associations between the senators and thus provides an alternative
view of the voting patterns, which together with the
thresholding technique employed gives a measure of the stability of
such associations. Further, the joint estimation method
allows one to separately study the overall voting patterns and those
driven by specific issues. In our view, both sets of techniques
are useful, with dimension reduction providing a global perspective and
the Markov model revealing more nuanced patterns.

\begin{appendix}
\section*{Appendix: Asymptotic properties}\label{app}
In this section we study the asymptotic properties of the proposed
joint estimation method. Since the structure of the underlying network
only depends on the interaction effects, we focus on a variant of the
model without main effects. Specifically, we solve
%
\begin{eqnarray}
\label{objfun_joint_nomaineffect}\quad&& \max_{\{\MM{\Theta}^{(k)}\}_{k=1}^K}  \sum_{k=1}^K
\frac{1}{n_k} \sum_{i=1}^{n_k} \sum
_{j=1}^p \biggl[x_{i,j}^{(k)}
\biggl(\sum_{j' \neq j} \theta_{j,j'}^{(k)}
x_{i,j'}^{(k)} \biggr) \nonumber\\
&&\hspace*{103pt}{}- \log \biggl\{1 + \exp \biggl(\sum
_{j' \neq j} \theta_{j,j'}^{(k)}
x_{i,j'}^{(k)} \biggr) \biggr\} \biggr]
\\
&&\qquad{} - \lambda\sum_{j < j'} \sqrt{\sum
_{k=1}^K\bigl | \theta _{j,j'}^{(k)}\bigr|}.\nonumber
\end{eqnarray}
We will show that the estimator in criterion \eqref
{objfun_joint_nomaineffect} is consistent in terms of both parameter
estimation and model selection, when $p$ and $n$ go to infinity and the
tuning parameter $\lambda$ goes to zero at some appropriate rate. We
note that our results are pointwise rather than uniform in $\Theta$,
as is standard in the literature. Some interesting implications of
nonuniform bounds for sparse estimators in linear regression have
recently been discussed by \citet{LeebPoetscher2008,PoetscherLeeb2009},
although their conclusions do not apply to graphical models.

Before stating the main results, we introduce necessary notation and
regularity conditions. For each $k=1,\ldots, K$, denote $\VV{\theta
}^{(k)} = (\theta_{1,2}^{(k)},\ldots, \theta_{j,j'}^{(k)},\ldots,\break
\theta_{p-1,p}^{(k)})$ as a $p(p-1)/2$-dimensional vector, recording
all upper triangular elements in $\MM{\Theta}^{(k)}$. Let $\overline
{\VV{\theta}}^{(k)}$ be the true value of $\VV{\theta}^{(k)}$. Let
$\overline{\M{Q}}^{(k)}$ be the population Fisher information matrix
of the model in criterion \eqref{objfun_joint_nomaineffect} (see the
\hyperref[app]{Appendix} for a precise definition) and let $\mathcal{X}_{(i)}^{(k)}$
be a matrix with $p$ rows and $p(p-1)/2$ columns,
whose $(j,j')$th column is composed of zeros except for the $j$th
($j'$th) component being $x_{i,j'}$ ($x_{i,j}$). In addition, we define
$\overline{\M{U}}^{(k)} = E[\T{\MM{\mathcal{X}}_{(i)}^{(k)}} \M
{\mathcal{X}}_{(i)}^{(k)}]$. To index the zero and nonzero elements,
let $S_k = \{(j,j')\dvtx \theta_{j,j'}^{(k)} \neq0, 1 \le j < j' \le p\}$
and $S_k^c = \{(j,j')\dvtx \theta_{j,j'}^{(k)} = 0, 1 \le j < j' \le p\}$,
and let $S_{\cap} = \bigcap_{k=1}^K S_k$, $S_{\cup} = \bigcup_{k=1}^K S_k$. The cardinalities of $S_k$ and $S_{\cup}$ are denoted
by $q_k$ and $q$, respectively. For any matrix $\M{W}$ and subsets of
row and
column indices $\mathcal{U}$ and $\mathcal{V}$, let
$\M{W}_{\mathcal{U}, \mathcal{V}}$ be the matrix consisting of
rows $\mathcal{U}$ and columns $\mathcal{V}$ in $\M{W}$. Finally, let
$\Lambda_{\min}(\cdot)$ and $\Lambda_{\max}(\cdot)$ denote the
smallest and largest eigenvalue of a matrix, respectively.

The asymptotic properties of the joint estimation method rely on the
following regularity conditions:
\begin{enumerate}[(A)]
\item[(A)] Nonzero elements bounds: There exist positive constants
$\gamma_{\min}$ and $\gamma_{\max}$ such that:
\begin{enumerate}[(ii)]
\item[(i)] $\min_{1 \le k \le K} \min_{(j,j') \in S_k} |\overline
{\theta}_{j,j'}^{(k)}| \ge\gamma_{\min}$;
\item[(ii)] $\max_{1 \le k \le K} \max_{(j,j') \in S_k \setminus
S_{\cap}} |\overline{\theta}_{j,j'}^{(k)}| \le\gamma_{\max}$.
\end{enumerate}
\item[(B)] Dependency: There exist positive constants $\tau_{\min}$
and $\tau_{\max}$ such that for any $k=1,\ldots, K$,
%
\begin{equation}
\Lambda_{\min} \bigl(\overline{\M{Q}}_{S_k, S_k}^{(k)}
\bigr) \ge \tau_{\min
} \quad\mbox{and} \quad\Lambda_{\max} \bigl(\overline{
\M{U}}_{S_k,S_k}^{(k)}\bigr) \le\tau_{\max}.
\end{equation}
\item[(C)] Incoherence: There exists a constant $\tau\in(1-\sqrt
{\gamma_{\min}/4\gamma_{\max}},1)$ such that for any $k=1,\ldots, K$,
%
\begin{equation}
\bigl\| \overline{\M{Q}}_{S_k^c, S_k}^{(k)} \bigl(\overline{
\M{Q}}_{S_k,
S_k}^{(k)}\bigr)^{-1}\bigr\|_{\infty} \le1 -
\tau.
\end{equation}
\end{enumerate}
Condition (A) enforces a lower bound on the magnitudes of all nonzero
elements, as well as an upper bound on the magnitudes of those nonzero
elements associated with individual links. Conditions (B) and (C) bound
the amount of
dependence and the influence that the nonneighbors can have
on a given node, respectively. Conditions similar to (B) and (C) were also
assumed by \citet{MeinshausenBuhlmann06},
\citet{RavikumarWainwrightLafferty10},
\citet{PengWangZhouZhu09} and \citet{GuoLevinaMichailidisZhu10JOSE}.
Our conditions are most closely related to those of \citet
{GuoLevinaMichailidisZhu10JOSE}, but here they are extended to the
heterogenous data setting.

%
\begin{theorem}[(Parameter estimation)]\label{thm_estimation_consistency}
Suppose all regularity conditions hold. If the tuning
parameter $\lambda= C_{\lambda} \sqrt{(\log p)/ n}$ for some
constant $C_{\lambda} > (8-4\tau)\sqrt{\gamma_{\min}}/(1-\tau)$
and if $\min\{n/q^3, n_1/q_1^3,\ldots, n_K/q_K^3\} > (4/C) \log p$
for some constant $C=\min\{\tau_{\min}^2 \tau^2 / 288 (1-\tau)^2,
\tau_{\min}^2 \tau^2/72, \tau_{\min}\tau/48\}$, then there exists
a local maximizer of the criterion \eqref{objfun_joint_nomaineffect},
$\{\widehat{\MM{\theta}}^{(k)}\}_{k=1}^K$, such that, with
probability tending to 1,
%
\begin{equation}
\sum_{k=1}^K \bigl\| \widehat{\VV{
\theta}}^{(k)} - \overline{\VV{\theta }}^{(k)} \bigr\|_2
\le M \sqrt{\frac{q \log p}{n}},
\end{equation}
for some constant $M > (2K C_{\lambda} / \tau_{\min} \sqrt{\gamma
_{\min}}) (3-2\tau)/(2-\tau)$.
\end{theorem}

%
\begin{theorem}[(Structure selection)]\label{thm_modelselection_consistency}
Under conditions of Theorem~\ref{thm_estimation_consistency}, with probability tending to 1, the
maximizer $\{\widehat{\MM{\theta}}^{(k)}\}_{k=1}^K$ from Theorem~\ref{thm_estimation_consistency} satisfies
\begin{eqnarray*}
\widehat{\theta}_{j,j'}^{(k)} &\neq& 0\qquad\mbox{for all }
\bigl(j,j'\bigr) \in S_k, k=1,\ldots, K;
\\
\widehat{\theta}_{j,j'}^{(k)} &=& 0\qquad \mbox{for all }
\bigl(j,j'\bigr) \in S_k^c, k=1,\ldots, K.
\end{eqnarray*}
\end{theorem}

Theorems \ref{thm_estimation_consistency} and \ref
{thm_modelselection_consistency} establish the consistency in terms of
parameter estimation and
structure selection, respectively.

The main idea of the proofs is closely related to \citet
{GuoLevinaMichailidisZhu10JOSE}, and some strategies for dealing with
the joint estimation are borrowed from \citet{GuoLevinaMichailidisZhu09CGM}.
We introduce notation first. For the $k$th category, we define the
log-likelihood as
\[
l\bigl(\VV{\theta}^{(k)}\bigr)=\frac{1}{n_k} \sum
_{i=1}^{n_k} \sum_{j=1}^p
\biggl[x_{i,j}^{(k)} \biggl(\sum_{j' \neq j}
\theta_{j,j'}^{(k)} x_{i,j'}^{(k)}\biggr) -
\log\biggl\{1 + \exp\biggl(\sum_{j' \neq j}
\theta_{j,j'}^{(k)} x_{i,j'}^{(k)} \biggr)\biggr
\}\biggr],
\]
whose first derivative and second derivative are denoted by $\nabla
l(\VV{\theta}^{(k)})$ and $\nabla^2 l(\VV{\theta}^{(k)})$,
respectively. Note that $\nabla l(\VV{\theta}^{(k)})$ is a
$p(p-1)/2$-dimensional vector and $\nabla^2 l(\VV{\theta}^{(k)})$ is
a $p(p-1)/2 \times p(p-1)/2$ matrix.
Then, the population Fisher information matrix of
the model in \eqref{objfun_joint_nomaineffect} at
$\overline{\VV{\theta}}$ can be defined as
$\overline{\M{Q}}^{(k)} = -\F{E}[\nabla^2 l(\overline{\VV{\theta
}}^{(k)})]$, and its sample
counterpart is
$\widehat{\M{Q}}^{(k)} = -\nabla^2 l(\overline{\VV{\theta
}}^{(k)})$. We also write
$\widehat{\M{U}}^{(k)} = 1/n \sum_{i=1}^n \T{\MM{\mathcal{X}}_{(i)}^{(k)}}
\MM{\mathcal{X}}_{(i)}^{(k)}$ for the sample counterpart of $\overline
{\M{U}}^{(k)}$. Let $\underline{\VV{\theta}}^{(k)} = (\underline
{\theta}_{1,2}^{(k)},\ldots, \underline{\theta}_{j,j'}^{(k)},\ldots, \underline{\theta}_{p-1,p}^{(k)})$ be the same as $\VV
{\theta}^{(k)}$ except that all elements in $S_k^c$ are set to zero
and write $\VV{\delta}^{(k)} = \VV{\theta}^{(k)} - \overline{\VV
{\theta}}^{(k)}$ and $\underline{\VV{\delta}}^{(k)} = \underline{\VV
{\theta}}^{(k)} - \overline{\VV{\theta}}^{(k)}$. Finally, let
$\mathcal{W}$ be a subset of the index set $\{1, 2,\ldots,
p(p-1)/2\}$. For a $p(p-1)/2$-dimensional vector $\VV{\beta}$, we define
$\VV{\beta}_{\mathcal{W}}$ as the vector consisting of the
elements of $\VV{\beta}$ associated with
$\mathcal{W}$.

Next, we introduce a variant of criterion \eqref
{objfun_joint_nomaineffect} by restricting all true zeros in $\{\VV
{\theta}^{(k)}\}_{k=1}^{K}$ to be estimated as zero. Specifically, the
restricted criterion is formulated as follows:
%
\begin{equation}
\label{obj_restricted_Ising} \max_{\{\underline{\VV{\theta}}^{(k)}\}_{k=1}^K} \sum_{k=1}^K
l\bigl(\underline{\VV{\theta}}^{(k)}\bigr) - \lambda\sum
_{1 \le
j < j' \le p} \sqrt{\sum_{k=1}^K
\bigl|\underline{\theta}_{j,j'}^{(k)}\bigr|},
\end{equation}
and its maximizer is denoted by $\{\widehat{\underline{\VV{\theta
}}}^{(k)}\}_{k=1}^{K}$.
In addition, we consider the sample versions of regularity conditions
(B) and (C):
\begin{longlist}[(B$^{\prime}$)]
\item[(B$^{\prime}$)] \textit{Sample dependency}: There exist positive
constants $\tau_{\min}$ and $\tau_{\max}$ such that for any
$k=1,\ldots, K$,
%
\begin{equation}
\Lambda_{\min} \bigl(\widehat{\M{Q}}_{S_k, S_k}^{(k)}
\bigr) \ge \tau_{\min
} \quad\mbox{and} \quad\Lambda_{\max} \bigl(\widehat{
\M{U}}_{S_k,S_k}^{(k)}\bigr) \le\tau_{\max}.
\end{equation}
\item[(C$^{\prime}$)] \textit{Sample incoherence}: There exists a constant
$\tau\in(1-\sqrt{\gamma_{\min}/4\gamma_{\max}},1)$ such that for
any $k=1,\ldots, K$,
%
\begin{equation}
\bigl\|\widehat{\M{Q}}_{S_k^c, S_k}^{(k)} \bigl(\widehat{
\M{Q}}_{S_k,
S_k}^{(k)}\bigr)^{-1}\bigr\|_{\infty} \le1 -
\tau.
\end{equation}
\end{longlist}

For convenience of the readers, the proof of our main result is divided
into two parts: Part I presents the main idea of the proof by listing the
important propositions and the proofs of Theorems
\ref{thm_estimation_consistency} and
\ref{thm_modelselection_consistency}, whereas part II
contains additional technical details and proofs of propositions in
part I.

\subsection*{Part I: Propositions and proof of Theorems \protect\ref
{thm_estimation_consistency} and \protect\ref{thm_modelselection_consistency}}
The proof consists of the following
steps. Proposition~\ref{thm_restricted_consistency} shows that, under
sample regularity conditions (B$^{\prime}$) and
(C$^{\prime}$), the conclusions of Theorems \ref
{thm_estimation_consistency} and \ref{thm_modelselection_consistency}
hold for the
local maximizer of the restricted problem \eqref
{obj_restricted_Ising}. Next, Proposition~\ref{prop_equival_sample_population}
proves that the population regularity conditions (B) and (C) give rise
to their sample counterparts (B$^{\prime}$) and (C$^{\prime}$)
with probability tending to one, hence, the conclusions of
Proposition~\ref{thm_restricted_consistency} also hold with the
population regularity conditions. Last, we show that the local
maximizer of \eqref{obj_restricted_Ising} is also a local maximizer of
the original model \eqref{objfun_joint_nomaineffect}.
This is established via Proposition~\ref{KKT_conditions}, which sets
out the Karush--Kuhn--Tucker (KKT) conditions for the local maximizer of
criterion \eqref{objfun_joint_nomaineffect}, and Proposition~\ref
{prop_restricted_KKT}, which
shows that, with probability tending to one, the local maximizer of
\eqref{obj_restricted_Ising} satisfies these KKT conditions.


%
\begin{proposition}\label{thm_restricted_consistency}
Suppose condition \textup{(A)} and the sample conditions \textup{(B$^{\prime}$)} and
\textup{(C$^{\prime}$)} hold.
If the tuning parameter $\lambda= C_{\lambda} \sqrt{(\log p)/ n}$
for some constant $C_{\lambda} > (8-4\tau)\sqrt{\gamma_{\min
}}/(1-\tau)$ and
$q\sqrt{(\log p)/n} = o(1)$, then with probability tending to one,
there exists a local maximizer of the restricted criterion, $\{\widehat
{\underline{\VV{\theta}}}^{(k)}\}_{k=1}^K$, satisfying:
\begin{longlist}[(ii)]
\item[(i)] $\sum_{k=1}^K \| \widehat{\underline{\VV{\theta
}}}^{(k)} - \overline{\VV{\theta}}^{(k)} \|_2 \leq M \sqrt{q (\log p)/n}$
for some constant $M > (2K C_{\lambda} /\break \tau_{\min} \sqrt
{\gamma_{\min}}) [(3-2\tau)/(2-\tau)]$;
\item[(ii)] For each $k=1,\ldots, K$, $\widehat{\underline{\theta
}}_{j,j'}^{(k)}
\neq0$ for all $(j,j') \in S_k$ and $\widehat{\underline{\theta
}}_{j,j'}^{(k)} =0$
for all $(j,j') \in S_k^c$.
\end{longlist}
\end{proposition}


\begin{proposition}\label{prop_equival_sample_population}
Suppose the regularity conditions \textup{(B)} and \textup{(C)} hold, then for any
$\varepsilon> 0$, the following inequalities hold with probability
tending to one for all $k=1,\ldots, K$:
\begin{longlist}[(iii)]
\item[(i)] $\F{P} \{\Lambda_{\mathrm{min}} (\widehat{\M{Q}}_{S_k,S_k}^{(k)}) \le
\tau_{\mathrm{min}} - \varepsilon\} \le2 \exp\{-(\varepsilon^2/2) (n_k/q_k^2) +
2 \log q_k\}$;
\item[(ii)] $\F{P}\{\Lambda_{\max} (\widehat{\M{U}}_{S_k,S_k}^{(k)}) \ge
\tau_{\max} + \varepsilon\} \le2 \exp\{-(\varepsilon^2/2) (n_k/q_k^2) +
2 \log q_k\}$;
\item[(iii)] $\F{P}[\|\widehat{\M{Q}}_{S_k^c,S_k}^{(k)} (\widehat
{\M{Q}}_{S_k, S_k}^{(k)})^{-1}\|_{\infty} \ge1
- \tau/2] \le12 \exp(-C n_k/q_k^3 + 4\log p)$, for some constant $C
= \min\{\tau_{\mathrm{min}}^2
\tau^2 / 288 (1-\tau)^2, \tau_{\mathrm{min}}^2 \tau^2 / 72, \tau_{\mathrm{min}}
\tau/ 48\}$.
\end{longlist}
\end{proposition}


\begin{proposition}\label{KKT_conditions}
$\{\widehat{\VV{\theta}}\}_{k=1}^K$ is a local maximizer of problem
\eqref{objfun_joint_nomaineffect} if and only if the following
conditions hold for all $k=1,\ldots, K$:
%
\begin{eqnarray}
 \nabla_{j,j'} l \bigl(\widehat{\VV{
\theta}}^{(k)}\bigr) &=& \lambda \sgn\bigl(\widehat{\theta}_{j,j'}^{(k)}
\bigr) \Big/ \Biggl(\sum_{k=1}^K \bigl|\widehat {
\theta}_{j,j'}^{(k)}\bigr|\Biggr)^{1/2}\qquad \mbox{if }
\widehat{\theta}_{j,j'}^{(k)} \neq0;
\nonumber
\\[-8pt]
\\[-8pt]
\nonumber
\bigl|
\nabla_{j,j'} l \bigl(\widehat{\VV{\theta}}^{(k)}\bigr)\bigr| &<&
\lambda\Big/ \Biggl(\sum_{k=1}^K \bigl|\widehat{
\theta}_{j,j'}^{(k)}\bigr|\Biggr)^{1/2}  \qquad \mbox{if }
\widehat{\theta}_{j,j'}^{(k)}=0.
\end{eqnarray}
\end{proposition}


\begin{proposition}\label{prop_restricted_KKT} Under all conditions of
Proposition~\ref{thm_restricted_consistency},
with probability tending to one, we have, for each $k=1,\ldots,K$,
%
\begin{eqnarray}
\nabla_{j,j'} l \bigl(\widehat{\underline{\VV{
\theta}}}^{(k)}\bigr) &=& \lambda \sgn\bigl(\widehat{\underline{
\theta}}_{j,j'}^{(k)}\bigr) \Big/ \Biggl(\sum
_{k=1}^K \bigl| \widehat{\underline{\theta}}_{j,j'}^{(k)}\bigr|
\Biggr)^{1/2} \qquad \mbox{for all }\bigl(j,j'\bigr) \in
S_k;
\nonumber
\\[-8pt]
\\[-8pt]
\nonumber
\bigl|\nabla_{j,j'} l \bigl(\widehat{
\underline{\VV{\theta}}}^{(k)}\bigr)\bigr| &<& \lambda\Big/ \Biggl(\sum
_{k=1}^K \bigl| \widehat{\underline{\theta
}}_{j,j'}^{(k)}\bigr|\Biggr)^{1/2} \qquad \mbox{for all }
\bigl(j,j'\bigr) \in S_k^c.
\end{eqnarray}
\end{proposition}


\begin{pf*}{Proof of Theorems \protect\ref
{thm_estimation_consistency} and \protect\ref{thm_modelselection_consistency}}
The condition $\min\{n/q^3, n_1/q_1^3,\ldots,  n_K/\break q_K^3\} > (4/C)
\log p$ implies that, for each $k=1,\ldots, K$, we have $-C n_k /
q_k^3 + 4 \log p < 0$ and $-(\varepsilon^2/2) (n_k / q_k^2) + 2 \log q_k
< 0$ when $q_k$ is large enough. This condition also implies $q \sqrt
{(\log p)/n}=o(1)$. In addition, by Proposition~\ref
{prop_equival_sample_population}, the sample conditions (B$^{\prime}$)
and (C$^{\prime}$) hold with probability tending to one when
regularity conditions (B) and (C) hold. Therefore, by Proposition~\ref
{thm_restricted_consistency}, with probability tending to one, the
solution of the restricted problem $\{\widehat{\underline{\VV{\theta
}}}^{(k)}\}_{k=1}^K$ satisfies both parameter estimation consistency
and structure selection consistency. On the other hand, by Proposition~\ref{prop_restricted_KKT}, with probability tending to one, $\{
\widehat{\underline{\VV{\theta}}}^{(k)}\}_{k=1}^K$ also satisfies
the KKT conditions in Proposition~\ref{KKT_conditions}, thus, it is a
local maximizer of criterion \eqref{objfun_joint_nomaineffect}. This
proves Theorems \ref{thm_estimation_consistency} and \ref
{thm_modelselection_consistency}.
\end{pf*}

\subsection*{Part II: Proofs of propositions}
Before proving the propositions, we state a few lemmas which will be
used in the proofs. These lemmas are variants of Lemmas 1, 2 and 5 in
\citet{GuoLevinaMichailidisZhu10JOSE}, adapted to the settings of the
heterogenous model and, thus, the proofs are omitted here. Likewise,
the proof of Proposition~\ref{prop_equival_sample_population} is very
similar to the proof of Propositions 3 and 4 in \citet
{GuoLevinaMichailidisZhu10JOSE} and is omitted.

\begin{lemma}\label{lemma_Bound_first_derivative}
For each $k=1,\ldots, K$, with probability tending to 1, we have
$\|\nabla l
(\overline{\VV{\theta}}^{(k)})\|_{\infty} \le C_{\nabla} \sqrt
{(\log p)/ n}$ for
some constant $C_{\nabla} > 4$.
\end{lemma}

\begin{lemma}\label{lemma_Bound_residual}
If the sample dependency condition \textup{(B$^{\prime}$)} holds and\break
$q\sqrt{(\log p)/n} = o(1)$, then for any $\alpha_k \in[0, 1]$,
$k=1,\ldots, K$, the following inequality holds with probability
tending to 1:
%
\begin{equation}
-\sum_{k=1}^K \T{\VV{
\delta}_{S_k}^{(k)}} \bigl[\nabla^2 l \bigl(
\overline {\VV{\theta}}^{(k)} + \alpha_k \underline{\VV{
\delta}}^{(k)}\bigr)\bigr]_{S_k,S_k} \VV{\delta}_{S_k}^{(k)}
\ge \frac{1}{2} \tau_{\min} \sum_{k=1}^K
\bigl\|\underline{\VV{\delta }}^{(k)}\bigr\|_2^2.
\end{equation}
\end{lemma}

\begin{lemma}\label{lemma_bound_r}
Suppose the sample dependency condition \textup{(B)} holds. For any $\alpha_k
\in[0,1]$, $k=1,\ldots, K$, the following inequality holds with
probability tending to one:
%
\begin{equation}
\bigl\|\bigl[\nabla^2 l \bigl(\overline{\VV{\theta}}^{(k)} +
\alpha_k \underline {\VV{\delta}}^{(k)}\bigr) -
\nabla^2 l \bigl(\overline{\VV{\theta}}^{(k)}\bigr)\bigr]
\underline{\VV{\delta}}^{(k)}\bigr\| _{\infty} \le\tau_{\max} \bigl\|
\underline{\VV{\delta}}^{(k)}\bigr\|_2^2.
\end{equation}
\end{lemma}

\begin{pf*}{Proof of Proposition \protect\ref{thm_restricted_consistency}}
The main idea of the proof was first introduced in this context in
\citet{RothmanBickelLevinaZhu08} and has since been used by many
authors. Define
%
\begin{eqnarray}
\label{fun_G} &&\F{G}\bigl(\bigl\{\underline{\VV{\delta}}^{(k)}\bigr
\}_{k=1}^K\bigr)\nonumber\\
&&\qquad= -\sum_{k=1}^K
\bigl[l\bigl(\overline{\VV{\theta}}^{(k)} + \underline{\VV{
\delta}}^{(k)}\bigr) - l\bigl(\overline{\VV{\theta}}^{(k)}\bigr)
\bigr]
\\
&&\qquad\quad{}+ \lambda\sum_{1 \le j<j' \le p} \Biggl\{ \Biggl(\sum
_{k=1}^K \bigl|\overline{\theta}_{j,j'}^{(k)}
+ \underline{\delta}_{j,j'}^{(k)}\bigr|\Biggr)^{1/2} -
\Biggl(\sum_{k=1}^K \bigl|\overline {
\theta}_{j,j'}^{(k)}\bigr|\Biggr)^{1/2}\Biggr\}.\nonumber
\end{eqnarray}
It can be seen from \eqref{obj_restricted_Ising} that $\{\underline
{\widehat{\VV{\delta}}}^{(k)}\}_{k=1}^{K}$ minimizes $\F{G}(\{
\underline{\VV{\delta}}^{(k)}\}_{k=1}^K)$ and\break $\F{G}(\{\V{0}\}
_{k=1}^K) = 0$. Thus, we must have $\F{G}
(\{\widehat{\underline{\VV{\delta}}}^{(k)}\}_{k=1}^K) \le0$. If we
take a closed set
$\mathcal{A}$ which contains $\{\V{0}\}_{k=1}^K$ and show that $\F
{G}$ is
strictly positive everywhere on the boundary $\partial
\mathcal{A}$, then it implies that $\F{G}$ has a local minimum
inside $\mathcal{A}$, since $\F{G}$ is continuous and
$\F{G}(\{\V{0}\}_{k=1}^K)=0$. Specifically, we define\break
$\mathcal{A}=\{\{\underline{\VV{\delta}}^{(k)}\}_{k=1}^K\dvtx \sum_{k=1}^K \| \underline{\VV{\delta}}^{(k)} \|_2 \le Ma_n \}$,
with boundary
$\partial\mathcal{A}= \{\{\underline{\VV{\delta}}^{(k)}\}_{k=1}^K\dvtx\break \sum_{k=1}^K \| \underline{\VV{\delta}}^{(k)} \|_2 = Ma_n \}$, for
some constant $M > (2K C_{\lambda} / \tau_{\min} \sqrt{\gamma
_{\min}}) [(3-2\tau)/\break(2-\tau)]$ and $a_n = \sqrt{q (\log p) / n}$.
For any $\{\underline{\VV{\delta}}^{(k)}\}_{k=1}^K \in\partial
\mathcal{A}$, the Taylor
series expansion gives $\F{G}(\{\underline{\VV{\delta}}^{(k)}\}
_{k=1}^K) = I_1 + I_2 + I_3$,
where
%
\begin{eqnarray}
I_1 &=& -\sum_{k=1}^K
\bigl[\nabla l\bigl(\overline{\VV{\theta }}^{(k)}\bigr)
\bigr]_{S_k}^{\mathsf{T}} \VV{\delta}_{S_k}^{(k)},
\nonumber
\\
\qquad I_2 &=& -\sum_{k=1}^K \T{\VV{
\delta}_{S_k}^{(k)}} \bigl[\nabla^2 l\bigl(
\overline{\VV{\theta}}^{(k)} + \alpha_k \underline{\VV{
\delta}}^{(k)}\bigr)\bigr]_{S_k,S_k} \VV{\delta }_{S_k}^{(k)}\qquad
\mbox{for some } \alpha_k \in[0,1],\hspace*{-35pt}
\\
I_3 &=& \lambda\sum_{(j,j') \in S_{\cup}} \Biggl\{
\Biggl(\sum_{k=1}^K \bigl|\overline{
\theta}_{j,j'}^{(k)} + \underline{\delta}_{j,j'}^{(k)}\bigr|
\Biggr)^{1/2} - \Biggl(\sum_{k=1}^K
\bigl|\overline {\theta}_{j,j'}^{(k)}\bigr|\Biggr)^{1/2}\Biggr\}.\nonumber
\end{eqnarray}
Since $C_{\lambda} > (8-4\tau)\sqrt{\gamma_{\min}}/(1-\tau)$, we
have $[(1-\tau)/(2-\tau)]
C_{\lambda}/\sqrt{\gamma_{\min}} > 4$. By Lemma~\ref
{lemma_Bound_first_derivative},
%
\begin{eqnarray}
|I_1| &\le&\sum_{k=1}^K \bigl\|
\bigl[\nabla l\bigl(\overline{\VV{\theta }}^{(k)}\bigr)
\bigr]_{S_k}\bigr\|_{\infty} \bigl\|\VV{\delta}_{S_k}^{(k)}
\bigr\|_1
\nonumber
\\[-8pt]
\\[-8pt]
\nonumber
&\le& \bigl[(1-\tau)C_{\lambda}M\gamma_{\min}^{-1/2}
/ (2-\tau)\bigr] (q \log p)/n.
\end{eqnarray}
In addition, by condition $q \sqrt{(\log p) / n}=o(1)$, Lemma~\ref
{lemma_Bound_residual} holds and, thus,
%
\begin{equation}
I_2 \ge(\tau_{\min}/2) \sum_{k=1}^K
\bigl\|\underline{\VV{\delta }}^{(k)}\bigr\|_2^2 \ge\bigl[
\tau_{\min}/(2K)\bigr] M^2 q (\log p) / n.
\end{equation}
Finally, by the triangular inequality and regularity condition (A),
%
\begin{eqnarray}
|I_3| &\le& \lambda\sum_{(j,j') \in S_{\cup}}\sum
_{k=1}^K \frac
{||\overline{\theta}_{j,j'}^{(k)} +
\underline{\delta}_{j,j'}^{(k)}| - |\overline{\theta
}_{j,j'}^{(k)}||}{(\sum_{k=1}^K |\overline{\theta}_{j,j'}^{(k)} +
\underline{\delta}_{j,j'}^{(k)}|)^{1/2} + (\sum_{k=1}^K |\overline
{\theta}_{j,j'}^{(k)}|)^{1/2}}
\nonumber
\\
&\le& \bigl(\lambda\gamma_{\min}^{-1/2}\bigr) \sum
_{k=1}^K \sum_{(j,j') \in
S_{\cup}} \bigl|
\underline{\delta}_{j,j'}^{(k)}\bigr| \le\bigl(\lambda
q^{1/2} \gamma_{\min}^{-1/2}\bigr) \sum
_{k=1}^K\bigl \|\underline{\VV{\delta }}^{(k)}
\bigr\|_2
\\
&\le& \bigl(M C_{\lambda} \gamma_{\min}^{-1/2}\bigr) \bigl
\{q (\log p)/n\bigr\}.\nonumber
\end{eqnarray}
Then we have
%
\begin{equation}\qquad
\F{G}\bigl(\bigl\{\underline{\VV{\delta}}^{(k)}\bigr
\}_{k=1}^K\bigr) \ge M^2 \frac{q
\log p}{n}
\biggl(\frac{\tau_{\min}}{2K} - \frac{(1-\tau) C_{\lambda
}}{(2-\tau)M\gamma_{\min}^{1/2}} - \frac{C_{\lambda}}{M\gamma_{\min}^{1/2}} \biggr) > 0.
\end{equation}
The last inequality uses the condition $M > (2K C_{\lambda} / \tau
_{\min} \sqrt{\gamma_{\min}}) [(3-2\tau)/(2-\tau)]$. Therefore,
with probability
tending to 1, we have $\sum_{k=1}^K \| \widehat{\underline{\VV
{\theta}}}^{(k)} - \overline{\VV{\theta}}^{(k)} \|_2 \leq M \sqrt{q
(\log p) /n}$, and consequently claim (i) in Proposition~\ref
{thm_restricted_consistency} holds.

On the other hand, by the definition of $\widehat{\underline{\VV
{\theta}}}^{(k)}$, we have $\widehat{\underline{\theta}}_{j,j'}^{(k)}=0$
for all $(j,j') \in S_k^c$. By regularity condition (A) and
Proposition~\ref{thm_restricted_consistency}(i), for any $(j,j') \in
S_k$, $k=1,\ldots,K$, we have $|\widehat{\underline{\theta
}}_{j,j'}^{(k)}| \ge|\overline{\theta}_{j,j'}^{(k)}| - |\widehat
{\underline{\theta}}_{j,j'}^{(k)} - \overline{\theta}_{j,j'}^{(k)}|
\ge\gamma_{\min}/2 > 0$, when $n$ is large enough.
\end{pf*}

\begin{pf*}{Proof of Proposition \protect\ref{prop_restricted_KKT}}
By Proposition~\ref{thm_restricted_consistency}, with
probability tending to one, we have $\widehat{\underline{\theta
}}_{j,j'} \neq0$ for all $(j,j') \in S_k$.
Since $\{\widehat{\underline{\VV{\theta}}}^{(k)}\}_{k=1}^K$ is a
local maximizer of the restricted problem \eqref
{obj_restricted_Ising}, with
probability tending to one, $\nabla_{j,j'} l (\widehat{\underline{\VV
{\theta}}}^{(k)}) = \lambda\sgn(\widehat{\underline{\theta
}}_{j,j'}^{(k)}) / (\sum_{k=1}^K | \widehat{\underline{\theta
}}_{j,j'}^{(k)}|)^{1/2}$, for all $(j,j') \in S_k$.

To show the second claim, we apply the mean value theorem and write
$\nabla
l(\widehat{\underline{\VV{\theta}}}^{(k)}) = \nabla
l(\overline{\VV{\theta}}^{(k)}) + \V{r}^{(k)} -
\widehat{\M{Q}}^{(k)} \underline{\widehat{\VV{\delta}}}^{(k)}$,
where $\V{r}^{(k)} =
\{\nabla^2 l(\overline{\VV{\theta}}^{(k)}+\alpha_k
\widehat{\underline{\VV{\delta}}}^{(k)}) -  \nabla^2
l(\overline{\VV{\theta}}^{(k)})\}
\widehat{\underline{\VV{\delta}}}^{(k)}$. After some
simplifications, we have
%
\begin{eqnarray}\qquad
\bigl[\nabla l\bigl(\widehat{\underline{\VV{\theta}}}^{(k)}\bigr)
\bigr]_{S_k^c} &=& \bigl[\nabla l\bigl(\overline{\VV{\theta}}^{(k)}
\bigr)\bigr]_{S_k^c} + \V{r}_{S_k^c}^{(k)}
\nonumber
\\[-8pt]
\\[-8pt]
\nonumber
&&{}- \bigl[
\widehat{\M{Q}}_{S_k^c,S_k}^{(k)} \bigl(\widehat{\M{Q}}_{S_k,S_k}^{(k)}
\bigr)^{-1}\bigr] \bigl\{\bigl[\nabla l\bigl(\overline{\VV{\theta
}}^{(k)}\bigr)\bigr]_{S_k} + \V{r}_{S_k}^{(k)}
- \bigl[\nabla l\bigl(\widehat{\underline {\VV{\theta}}}^{(k)}\bigr)
\bigr]_{S_k}\bigr\}
\end{eqnarray}
and, thus,
%
\begin{eqnarray}
&&\bigl\|\bigl[\nabla l\bigl(\widehat{\underline{\VV{\theta}}}^{(k)}\bigr)
\bigr]_{S_k^c}\bigr\|_{\infty}\nonumber \\
&&\qquad\le \bigl\|\bigl[\nabla l\bigl(\overline{\VV{
\theta}}^{(k)}\bigr)\bigr]_{S_k^c}\bigr\|_{\infty} + \bigl\|
\V{r}_{S_k^c}^{(k)}\bigr\|_{\infty}
\nonumber
\\
&&\qquad\quad{}+ \bigl\|\widehat{\M{Q}}_{S_k^c, S_k}^{(k)} \bigl(\widehat{
\M{Q}}_{S_k,
S_k}^{(k)}\bigr)^{-1}\bigr\|_{\infty}\nonumber\\
&&\qquad\quad{}\times \bigl\{
\bigl\|\bigl[\nabla l\bigl(\overline{\VV{\theta}}^{(k)}\bigr)
\bigr]_{S_k}\bigr\|_{\infty} + \bigl\|\V{r}_{S_k}^{(k)}
\bigr\|_{\infty} + \bigl\|\bigl[\nabla l\bigl(\widehat{\underline{\VV{
\theta}}}^{(k)}\bigr)\bigr]_{S_k}\bigr\|_{\infty}\bigr\}
\nonumber
\\[-8pt]
\\[-8pt]
\nonumber
&&\qquad\le (2-\tau)\bigl \|\nabla l\bigl(\overline{\VV{\theta}}^{(k)}\bigr)
\bigr\|_{\infty} + (2- \tau)\bigl\|\V{r}^{(k)}\bigr\|_{\infty} + (1-\tau) \bigl\|
\bigl[\nabla l\bigl(\widehat{\underline{\VV{\theta}}}^{(k)}\bigr)
\bigr]_{S_k}\bigr\|_{\infty}
\\
&&\qquad\le \bigl[(1-\tau)C_{\lambda}/\sqrt{\gamma_{\min}}\bigr] \sqrt{(
\log p)/n} + (2-\tau) \tau_{\max} M^2 q (\log p) / n
\nonumber
\\
&&\qquad\quad{}+ (1-\tau) \lambda\Big/ \min_{(j,j') \in S_k} \Biggl[\sum
_{k=1}^K \bigl|\widehat{\underline{\theta}}_{j,j'}\bigr|
\Biggr]^{1/2}
\nonumber
\\
&&\qquad\le \bigl[2(1-\tau)/\sqrt{\gamma_{\min}}\bigr] \lambda+
o_p(\lambda).\nonumber
\end{eqnarray}
On the other hand, $\lambda/ [\sum_{k=1}^K
|\widehat{\underline{\theta}}_{j,j'}^{(k)}|]^{1/2} = +\infty$ when
$(j,j') \in S_{\cup}^c$. Otherwise, if
$(j,j') \in S_{\cup} \setminus S_k$, then
\begin{eqnarray*}
\lambda\Big/ \Biggl(\sum_{k=1}^K \bigl|\widehat{
\underline{\theta}}_{j,j'}\bigr|\Biggr)^{1/2} &\ge&\lambda\Big/ \Biggl\{\sum
_{k=1}^K \bigl|\widehat{\underline{
\theta}}_{j,j'} - \overline{\theta}_{j,j'}\bigr| + \bigl|\overline{
\theta}_{j,j'}\bigr|\Biggr\}^{1/2}\\
& \ge&\lambda/ \sqrt{\gamma
_{\max}} \ge(2-2\tau) \lambda/ \sqrt{\gamma_{\min}}. %
\end{eqnarray*}
Thus, for any $(j,j') \in S_k^c$ ($k=1,\ldots,K$), we have
%
\begin{eqnarray}
\bigl|\nabla_{j,j'} l\bigl(\widehat{\underline{\VV{\theta}}}^{(k)}
\bigr)\bigr| &\le& \max_{1 \le k \le K}\max_{(j,j') \in S_k^c}\bigl |
\nabla_{j,j'} l\bigl(\widehat{\underline{\VV{\theta}}}^{(k)}
\bigr)\bigr|
\nonumber
\\[-8pt]
\\[-8pt]
\nonumber
&< &\min_{1 \le k \le K} \min_{(j,j') \in S_k^c} \lambda\Big/ \sqrt
{\sum_{k=1}^K \bigl|\widehat{
\underline{\theta}}_{j,j'}^{(k)}\bigr|} \le \lambda\Big/ \sqrt{
\sum_{k=1}^K \bigl|\widehat{\underline{\theta
}}_{j,j'}^{(k)}\bigr|}.
\end{eqnarray}
\upqed\end{pf*}
\end{appendix}

%





\printaddresses

\begin{thebibliography}{39}

\bibitem[\protect\citeauthoryear{Airoldi}{2007}]{Airoldi07}
%
\begin{barticle}[pbm]
\bauthor{\bsnm{Airoldi},~\bfnm{Edoardo~M.}\binits{E.~M.}}
(\byear{2007}).
\btitle{Getting started in probabilistic graphical models}.
\bjournal{PLoS Comput. Biol.}
\bvolume{3}
\bpages{e252}.
\bid{doi={10.1371/journal.pcbi.0030252}, issn={1553-7358},
pii={07-PLCB-MI-0342}, pmcid={2134967}, pmid={18069887}}
\end{barticle}
%
\bptok{imsref}%
\endbibitem

\bibitem[\protect\citeauthoryear{Anandkumar et~al.}{2012}]{Anandkumar12}
%
\begin{barticle}[mr]
\bauthor{\bsnm{Anandkumar},~\bfnm{Animashree}\binits{A.}},
\bauthor{\bsnm{Tan},~\bfnm{Vincent~Y.~F.}\binits{V.~Y.~F.}},
\bauthor{\bsnm{Huang},~\bfnm{Furong}\binits{F.}} \AND
\bauthor{\bsnm{Willsky},~\bfnm{Alan~S.}\binits{A.~S.}}
(\byear{2012}).
\btitle{High-dimensional structure estimation in {I}sing models: Local
separation criterion}.
\bjournal{Ann. Statist.}
\bvolume{40}
\bpages{1346--1375}.
\bid{doi={10.1214/12-AOS1009}, issn={0090-5364}, mr={3015028}}
\end{barticle}
%
\bptok{imsref}%
\endbibitem

\bibitem[\protect\citeauthoryear{Banerjee, El~Ghaoui and
d'Aspremont}{2008}]{BanerjeeGhaouidAspremont08}
%
\begin{barticle}[mr]
\bauthor{\bsnm{Banerjee},~\bfnm{Onureena}\binits{O.}},
\bauthor{\bsnm{El Ghaoui},~\bfnm{Laurent}\binits{L.}} \AND
\bauthor{\bsnm{d'Aspremont},~\bfnm{Alexandre}\binits{A.}}
(\byear{2008}).
\btitle{Model selection through sparse maximum likelihood estimation
for multivariate {G}aussian or binary data}.
\bjournal{J. Mach. Learn. Res.}
\bvolume{9}
\bpages{485--516}.
\bid{issn={1532-4435}, mr={2417243}}
\end{barticle}
%
\bptok{imsref}%
\endbibitem

\bibitem[\protect\citeauthoryear{Barab\'asi and
Albert}{1999}]{BarabasiAlbert99}
%
\begin{barticle}[mr]
\bauthor{\bsnm{Barab\'asi},~\bfnm{Albert-L\'aszl\'o}\binits
{A.-L.}} \AND
\bauthor{\bsnm{Albert},~\bfnm{R\'eka}\binits{R.}}
(\byear{1999}).
\btitle{Emergence of scaling in random networks}.
\bjournal{Science}
\bvolume{286}
\bpages{509--512}.
\bid{doi={10.1126/science.286.5439.509}, issn={0036-8075}, mr={2091634}}
\end{barticle}
%
\bptok{imsref}%
\endbibitem

\bibitem[\protect\citeauthoryear{Besag}{1986}]{besag86}
%
\begin{barticle}[mr]
\bauthor{\bsnm{Besag},~\bfnm{Julian}\binits{J.}}
(\byear{1986}).
\btitle{On the statistical analysis of dirty pictures}.
\bjournal{J. R. Stat. Soc. Ser. B Stat. Methodol.}
\bvolume{48}
\bpages{259--302}.
\bid{issn={0035-9246}, mr={0876840}}
\end{barticle}
%
\bptok{imsref}%
\endbibitem

\bibitem[\protect\citeauthoryear{Clinton, Jackman and
Rivers}{2004}]{ClintonJackmanRivers04}
%
\begin{barticle}[author]
\bauthor{\bsnm{Clinton},~\bfnm{J.}\binits{J.}},
\bauthor{\bsnm{Jackman},~\bfnm{S.}\binits{S.}} \AND
\bauthor{\bsnm{Rivers},~\bfnm{D.}\binits{D.}}
(\byear{2004}).
\btitle{The statistical analysis of roll call data}.
\bjournal{American Political Science Review}
\bvolume{98}
\bpages{355--370}.
\end{barticle}
%
\bptok{imsref}%
\endbibitem

\bibitem[\protect\citeauthoryear{Danaher, Wang and Witten}{2011}]{Danaher}
%
\begin{bmisc}[author]
\bauthor{\bsnm{Danaher},~\bfnm{P.}\binits{P.}},
\bauthor{\bsnm{Wang},~\bfnm{P.}\binits{P.}} \AND
\bauthor{\bsnm{Witten},~\bfnm{D.~M.}\binits{D.~M.}}
(\byear{2011}).
\bhowpublished{The joint graphical lasso for inverse covariance estimation
across multiple classes.
Available at \arxivurl{arXiv:1111.0324}.}
\end{bmisc}
%
\bptok{imsref}%
\endbibitem

\bibitem[\protect\citeauthoryear{de~Leeuw}{2006}]{deLeeuw06}
%
\begin{bincollection}[author]
\bauthor{\bparticle{de} \bsnm{Leeuw},~\bfnm{J.}\binits{J.}}
(\byear{2006}).
\btitle{Principal component analysis of senate voting patterns}.
In \bbooktitle{Real Data Analysis}
(\beditor{\bfnm{S. S.}\binits{S. S.}~\bsnm{Sawilowski}}, ed.)
\bpages{405--411}.
\bpublisher{Information Age Publishing},
\blocation{Charlotte, NC}.
\end{bincollection}
%
\bptok{imsref}%
\endbibitem

\bibitem[\protect\citeauthoryear{Diaconis, Goel and
Holmes}{2008}]{DiaconisGoelHolmes08}
%
\begin{barticle}[mr]
\bauthor{\bsnm{Diaconis},~\bfnm{Persi}\binits{P.}},
\bauthor{\bsnm{Goel},~\bfnm{Sharad}\binits{S.}} \AND
\bauthor{\bsnm{Holmes},~\bfnm{Susan}\binits{S.}}
(\byear{2008}).
\btitle{Horseshoes in multidimensional scaling and local kernel methods}.
\bjournal{Ann. Appl. Stat.}
\bvolume{2}
\bpages{777--807}.
\bid{doi={10.1214/08-AOAS165}, issn={1932-6157}, mr={2516794}}
\end{barticle}
%
\bptok{imsref}%
\endbibitem

\bibitem[\protect\citeauthoryear{Enelow and Hinich}{1984}]{EnelowHinich84}
%
\begin{bbook}[author]
\bauthor{\bsnm{Enelow},~\bfnm{J.~M.}\binits{J.~M.}} \AND
\bauthor{\bsnm{Hinich},~\bfnm{M.~J.}\binits{M.~J.}}
(\byear{1984}).
\btitle{The Spatial Theory of Voting: An Introduction}.
\bpublisher{Cambridge Univ. Press},
\blocation{Cambridge}.
\end{bbook}
%
\bptok{imsref}%
\endbibitem

\bibitem[\protect\citeauthoryear{Gerrish}{2011}]{Gerrish2011}
%
\begin{binproceedings}[author]
\bauthor{\bsnm{Gerrish},~\bfnm{Sean~M.}\binits{S.~M.}}
(\byear{2011}).
\btitle{Predicting legislative roll calls from text}.
In \bbooktitle{Proc. 28th Internat. Conf. on Machine Learning
(ICML-11)}.
\bpublisher{Omnipress},
\blocation{Madison, WI}.
\end{binproceedings}
%
\bptok{imsref}%
\endbibitem





\bibitem[\protect\citeauthoryear{Guo
et~al.}{2009}]{GuoLevinaMichailidisZhu09jointIsingtechreport}
%
\begin{bmisc}[author]
\bauthor{\bsnm{Guo},~\bfnm{J.}\binits{J.}},
\bauthor{\bsnm{Levina},~\bfnm{E.}\binits{E.}},
\bauthor{\bsnm{Michailidis},~\bfnm{G.}\binits{G.}} \AND
\bauthor{\bsnm{Zhu},~\bfnm{J.}\binits{J.}}
(\byear{2009}).
\bhowpublished{Joint structure estimation of {M}arkov network.
Technical report,
Dept. Statistics, Univ. Michigan, Ann Arbor, MI}.
\end{bmisc}
%
\bptok{imsref}%
\endbibitem

\bibitem[\protect\citeauthoryear{Guo
et~al.}{2010}]{GuoLevinaMichailidisZhu10JOSE}
%
\begin{bmisc}[author]
\bauthor{\bsnm{Guo},~\bfnm{J.}\binits{J.}},
\bauthor{\bsnm{Levina},~\bfnm{E.}\binits{E.}},
\bauthor{\bsnm{Michailidis},~\bfnm{G.}\binits{G.}} \AND
\bauthor{\bsnm{Zhu},~\bfnm{J.}\binits{J.}}
(\byear{2010}).
\bhowpublished{Joint structure estimation for categorical {M}arkov networks.
Technical report,
Dept. Statistics, Univ. Michigan, Ann Arbor, MI}.
\end{bmisc}
%
\bptok{imsref}%
\endbibitem

\bibitem[\protect\citeauthoryear{Guo
et~al.}{2011}]{GuoLevinaMichailidisZhu09CGM}
%
\begin{barticle}[mr]
\bauthor{\bsnm{Guo},~\bfnm{Jian}\binits{J.}},
\bauthor{\bsnm{Levina},~\bfnm{Elizaveta}\binits{E.}},
\bauthor{\bsnm{Michailidis},~\bfnm{George}\binits{G.}} \AND
\bauthor{\bsnm{Zhu},~\bfnm{Ji}\binits{J.}}
(\byear{2011}).
\btitle{Joint estimation of multiple graphical models}.
\bjournal{Biometrika}
\bvolume{98}
\bpages{1--15}.
\bid{doi={10.1093/biomet/asq060}, issn={0006-3444}, mr={2804206}}
\end{barticle}
%
\bptok{imsref}%
\endbibitem

\bibitem[\protect\citeauthoryear{Han}{2007}]{Han2007}
%
\begin{barticle}[author]
\bauthor{\bsnm{Han},~\bfnm{J~H.}\binits{J.~H.}}
(\byear{2007}).
\btitle{Analysing roll calls of the European Parliament: A Bayesian
application}.
\bjournal{European Union Politics}
\bvolume{8}
\bpages{479--507}.
\end{barticle}
%
\bptok{imsref}%
\endbibitem

\bibitem[\protect\citeauthoryear{Hara and Washio}{2013}]{Hara201323}
%
\begin{barticle}[author]
\bauthor{\bsnm{Hara},~\bfnm{Satoshi}\binits{S.}} \AND
\bauthor{\bsnm{Washio},~\bfnm{Takashi}\binits{T.}}
(\byear{2013}).
\btitle{Learning a common substructure of multiple graphical Gaussian models}.
\bjournal{Neural Networks}
\bvolume{38}
\bpages{23--38}.
\end{barticle}
%
\bptok{imsref}%
\endbibitem

\bibitem[\protect\citeauthoryear{H\"ofling and
Tibshirani}{2009}]{HoeflingTibshirani09}
%
\begin{barticle}[mr]
\bauthor{\bsnm{H\"ofling},~\bfnm{Holger}\binits{H.}} \AND
\bauthor{\bsnm{Tibshirani},~\bfnm{Robert}\binits{R.}}
(\byear{2009}).
\btitle{Estimation of sparse binary pairwise {M}arkov networks using
pseudo-likelihoods}.
\bjournal{J. Mach. Learn. Res.}
\bvolume{10}
\bpages{883--906}.
\bid{issn={1532-4435}, mr={2505138}}
\end{barticle}
%
\bptok{imsref}%
\endbibitem

\bibitem[\protect\citeauthoryear{H{\o}jsgaard}{2004}]{Hojsgaard04}
%
\begin{barticle}[mr]
\bauthor{\bsnm{H{\o}jsgaard},~\bfnm{S{\o}ren}\binits{S.}}
(\byear{2004}).
\btitle{Statistical inference in context specific interaction models
for contingency tables}.
\bjournal{Scand. J. Stat.}
\bvolume{31}
\bpages{143--158}.
\bid{doi={10.1111/j.1467-9469.2004.00378.x}, issn={0303-6898}, mr={2042604}}
\end{barticle}
%
\bptok{imsref}%
\endbibitem

\bibitem[\protect\citeauthoryear{Jung et~al.}{1996}]{JungParkChoiKim96}
%
\begin{binproceedings}[author]
\bauthor{\bsnm{Jung},~\bfnm{S.~Y.}\binits{S.~Y.}},
\bauthor{\bsnm{Park},~\bfnm{Y.~C.}\binits{Y.~C.}},
\bauthor{\bsnm{Choi},~\bfnm{K.~S.}\binits{K.~S.}} \AND
\bauthor{\bsnm{Kim},~\bfnm{Y.}\binits{Y.}}
(\byear{1996}).
\btitle{Markov random field based English part-of-speech tagging system}.
In \bbooktitle{Proceedings of the 16th Conference on Computational Linguistics}
\bpages{236--242}.
\bpublisher{Association for Computational Linguistics},
\blocation{Stroudsburg, PA}.
\end{binproceedings}
%
\bptok{imsref}%
\endbibitem

\bibitem[\protect\citeauthoryear{Kolar and Xing}{2008}]{KolarXing08}
%
\begin{bmisc}[author]
\bauthor{\bsnm{Kolar},~\bfnm{M.}\binits{M.}} \AND
\bauthor{\bsnm{Xing},~\bfnm{E.~P.}\binits{E.~P.}}
(\byear{2008}).
\bhowpublished{Improved estimation of high-dimensional {I}sing models.
Available at \arxivurl{arXiv:0811.1239}.}
\end{bmisc}
%
\bptok{imsref}%
\endbibitem

\bibitem[\protect\citeauthoryear{Leeb and P\"otscher}{2008}]{LeebPoetscher2008}
%
\begin{barticle}[mr]
\bauthor{\bsnm{Leeb},~\bfnm{Hannes}\binits{H.}} \AND
\bauthor{\bsnm{P\"otscher},~\bfnm{Benedikt~M.}\binits{B.~M.}}
(\byear{2008}).
\btitle{Sparse estimators and the oracle property, or the return of
{H}odges' estimator}.
\bjournal{J. Econometrics}
\bvolume{142}
\bpages{201--211}.
\bid{doi={10.1016/j.jeconom.2007.05.017}, issn={0304-4076}, mr={2394290}}
\end{barticle}
%
\bptok{imsref}%
\endbibitem

\bibitem[\protect\citeauthoryear{Li}{2001}]{Li01}
%
\begin{bbook}[author]
\bauthor{\bsnm{Li},~\bfnm{S.~Z.}\binits{S.~Z.}}
(\byear{2001}).
\btitle{Markov Random Field Modeling in Image Analysis}.
\bpublisher{Springer},
\blocation{New York}.
\end{bbook}
%
\bptok{imsref}%
\endbibitem

\bibitem[\protect\citeauthoryear{Li and Gui}{2006}]{LiGui06}
%
\begin{barticle}[pbm]
\bauthor{\bsnm{Li},~\bfnm{Hongzhe}\binits{H.}} \AND
\bauthor{\bsnm{Gui},~\bfnm{Jiang}\binits{J.}}
(\byear{2006}).
\btitle{Gradient directed regularization for sparse Gaussian
concentration graphs, with applications to inference of genetic networks}.
\bjournal{Biostatistics}
\bvolume{7}
\bpages{302--317}.
\bid{doi={10.1093/biostatistics/kxj008}, issn={1465-4644},
pii={kxj008}, pmid={16326758}}
\end{barticle}
%
\bptok{imsref}%
\endbibitem

\bibitem[\protect\citeauthoryear{Matthews and Stimson}{1975}]{MatthewsStimson75}
%
\begin{bbook}[author]
\bauthor{\bsnm{Matthews},~\bfnm{D.~R.}\binits{D.~R.}} \AND
\bauthor{\bsnm{Stimson},~\bfnm{J.~A.}\binits{J.~A.}}
(\byear{1975}).
\btitle{Yeas and Nays: Normal Decision-Making in the {U.S.} {H}ouse of
{R}epresentatives}.
\bpublisher{Wiley},
\blocation{New York}.
\end{bbook}
%
\bptok{imsref}%
\endbibitem

\bibitem[\protect\citeauthoryear{Meinshausen and B\"uhlmann}{2006}]{MeinshausenBuhlmann06}
%
\begin{barticle}[mr]
\bauthor{\bsnm{Meinshausen},~\bfnm{Nicolai}\binits{N.}} \AND
\bauthor{\bsnm{B\"uhlmann},~\bfnm{Peter}\binits{P.}}
(\byear{2006}).
\btitle{High-dimensional graphs and variable selection with the lasso}.
\bjournal{Ann. Statist.}
\bvolume{34}
\bpages{1436--1462}.
\bid{doi={10.1214/009053606000000281}, issn={0090-5364}, mr={2278363}}
\end{barticle}
%
\bptok{imsref}%
\endbibitem

\bibitem[\protect\citeauthoryear{Meinshausen and B\"uhlmann}{2010}]{MeinshausenBuhlmann10}
%
\begin{barticle}[mr]
\bauthor{\bsnm{Meinshausen},~\bfnm{Nicolai}\binits{N.}} \AND
\bauthor{\bsnm{B\"uhlmann},~\bfnm{Peter}\binits{P.}}
(\byear{2010}).
\btitle{Stability selection}.
\bjournal{J. R. Stat. Soc. Ser. B Stat. Methodol.}
\bvolume{72}
\bpages{417--473}.
\bid{doi={10.1111/j.1467-9868.2010.00740.x}, issn={1369-7412}, mr={2758523}}
\bptnote{check related}%
\end{barticle}
%
\bptok{imsref}%
\endbibitem

\bibitem[\protect\citeauthoryear{Morton}{1999}]{Morton99}
%
\begin{bbook}[author]
\bauthor{\bsnm{Morton},~\bfnm{R.~B.}\binits{R.~B.}}
(\byear{1999}).
\btitle{Methods and Models: A Guide to the Empirical Analysis of
Formal Models in Political Science}.
\bpublisher{Cambridge Univ. Press},
\blocation{Cambridge}.
\end{bbook}
%
\bptok{imsref}%
\endbibitem

\bibitem[\protect\citeauthoryear{Peng, Zhou and
Zhu}{2009}]{PengWangZhouZhu09}
%
\begin{barticle}[mr]
\bauthor{\bsnm{Peng},~\bfnm{Jie}\binits{J.}},
\bauthor{\bsnm{Zhou},~\bfnm{Nengfeng}\binits{N.}} \AND
\bauthor{\bsnm{Zhu},~\bfnm{Ji}\binits{J.}}
(\byear{2009}).
\btitle{Partial correlation estimation by joint sparse regression models}.
\bjournal{J. Amer. Statist. Assoc.}
\bvolume{104}
\bpages{735--746}.
\bid{doi={10.1198/jasa.2009.0126}, issn={0162-1459}, mr={2541591}}
\end{barticle}
%
\bptok{imsref}%
\endbibitem

\bibitem[\protect\citeauthoryear{Poole and
Rosenthal}{1997}]{PooleRosenthal97}
%
\begin{bbook}[author]
\bauthor{\bsnm{Poole},~\bfnm{K.~T.}\binits{K.~T.}} \AND
\bauthor{\bsnm{Rosenthal},~\bfnm{H.}\binits{H.}}
(\byear{1997}).
\btitle{Congress: A Political-Economic History of Roll-Call Voting}.
\bpublisher{Oxford Univ. Press},
\blocation{Oxford}.
\end{bbook}
%
\bptok{imsref}%
\endbibitem

\bibitem[\protect\citeauthoryear{P\"otscher and
Leeb}{2009}]{PoetscherLeeb2009}
%
\begin{barticle}[mr]
\bauthor{\bsnm{P\"otscher},~\bfnm{Benedikt~M.}\binits{B.~M.}}
\AND
\bauthor{\bsnm{Leeb},~\bfnm{Hannes}\binits{H.}}
(\byear{2009}).
\btitle{On the distribution of penalized maximum likelihood
estimators: The LASSO, SCAD, and thresholding}.
\bjournal{J. Multivariate Anal.}
\bvolume{100}
\bpages{2065--2082}.
\bid{doi={10.1016/j.jmva.2009.06.010}, issn={0047-259X}, mr={2543087}}
\end{barticle}
%
\bptok{imsref}%
\endbibitem

\bibitem[\protect\citeauthoryear{Ravikumar, Wainwright and
Lafferty}{2010}]{RavikumarWainwrightLafferty10}
%
\begin{barticle}[mr]
\bauthor{\bsnm{Ravikumar},~\bfnm{Pradeep}\binits{P.}},
\bauthor{\bsnm{Wainwright},~\bfnm{Martin~J.}\binits{M.~J.}} \AND
\bauthor{\bsnm{Lafferty},~\bfnm{John~D.}\binits{J.~D.}}
(\byear{2010}).
\btitle{High-dimensional {I}sing model selection using {$\ell_1$}-regularized
logistic regression}.
\bjournal{Ann. Statist.}
\bvolume{38}
\bpages{1287--1319}.
\bid{doi={10.1214/09-AOS691}, issn={0090-5364}, mr={2662343}}
\end{barticle}
%
\bptok{imsref}%
\endbibitem

\bibitem[\protect\citeauthoryear{Ravikumar
et~al.}{2011}]{RavikumarWainwrightRaskuttiYu08}
%
\begin{barticle}[mr]
\bauthor{\bsnm{Ravikumar},~\bfnm{Pradeep}\binits{P.}},
\bauthor{\bsnm{Wainwright},~\bfnm{Martin~J.}\binits{M.~J.}},
\bauthor{\bsnm{Raskutti},~\bfnm{Garvesh}\binits{G.}} \AND
\bauthor{\bsnm{Yu},~\bfnm{Bin}\binits{B.}}
(\byear{2011}).
\btitle{High-dimensional covariance estimation by minimizing {$\ell_1$}-penalized log-determinant divergence}.
\bjournal{Electron. J. Stat.}
\bvolume{5}
\bpages{935--980}.
\bid{doi={10.1214/11-EJS631}, issn={1935-7524}, mr={2836766}}
\end{barticle}
%
\bptok{imsref}%
\endbibitem

\bibitem[\protect\citeauthoryear{Rothman
et~al.}{2008}]{RothmanBickelLevinaZhu08}
%
\begin{barticle}[mr]
\bauthor{\bsnm{Rothman},~\bfnm{Adam~J.}\binits{A.~J.}},
\bauthor{\bsnm{Bickel},~\bfnm{Peter~J.}\binits{P.~J.}},
\bauthor{\bsnm{Levina},~\bfnm{Elizaveta}\binits{E.}} \AND
\bauthor{\bsnm{Zhu},~\bfnm{Ji}\binits{J.}}
(\byear{2008}).
\btitle{Sparse permutation invariant covariance estimation}.
\bjournal{Electron. J. Stat.}
\bvolume{2}
\bpages{494--515}.
\bid{doi={10.1214/08-EJS176}, issn={1935-7524}, mr={2417391}}
\end{barticle}
%
\bptok{imsref}%
\endbibitem

\bibitem[\protect\citeauthoryear{Wainwright and
Jordan}{2008}]{WainwrightJordan08}
%
\begin{barticle}[author]
\bauthor{\bsnm{Wainwright},~\bfnm{M.~J.}\binits{M.~J.}} \AND
\bauthor{\bsnm{Jordan},~\bfnm{M.~I.}\binits{M.~I.}}
(\byear{2008}).
\btitle{Graphical models, exponential families, and variational inference}.
\bjournal{Foundations and Trends in Machine Learning}
\bvolume{1}
\bpages{1--305}.
\end{barticle}
%
\bptok{imsref}%
\endbibitem

\bibitem[\protect\citeauthoryear{Xue, Zou and Cai}{2012}]{Xue12}
%
\begin{barticle}[mr]
\bauthor{\bsnm{Xue},~\bfnm{Lingzhou}\binits{L.}},
\bauthor{\bsnm{Zou},~\bfnm{Hui}\binits{H.}} \AND
\bauthor{\bsnm{Cai},~\bfnm{Tianxi}\binits{T.}}
(\byear{2012}).
\btitle{Nonconcave penalized composite conditional likelihood
estimation of sparse {I}sing models}.
\bjournal{Ann. Statist.}
\bvolume{40}
\bpages{1403--1429}.
\bid{doi={10.1214/12-AOS1017}, issn={0090-5364}, mr={3015030}}
\end{barticle}
%
\bptok{imsref}%
\endbibitem

\bibitem[\protect\citeauthoryear{Yang et~al.}{2012}]{Yang12}
%
\begin{bmisc}[author]
\bauthor{\bsnm{Yang},~\bfnm{Sen}\binits{S.}},
\bauthor{\bsnm{Pan},~\bfnm{Zhisong}\binits{Z.}},
\bauthor{\bsnm{Shen},~\bfnm{Xiaotong}\binits{X.}},
\bauthor{\bsnm{Wonka},~\bfnm{Peter}\binits{P.}} \AND
\bauthor{\bsnm{Ye},~\bfnm{Jieping}\binits{J.}}
(\byear{2012}).
\bhowpublished{Fused multiple graphical lasso.
Available at \arxivurl{arXiv:1209.2139}.}
\end{bmisc}
%
\bptok{imsref}%
\endbibitem

\bibitem[\protect\citeauthoryear{Yuan and Lin}{2007}]{YuanLin07}
%
\begin{barticle}[mr]
\bauthor{\bsnm{Yuan},~\bfnm{Ming}\binits{M.}} \AND
\bauthor{\bsnm{Lin},~\bfnm{Yi}\binits{Y.}}
(\byear{2007}).
\btitle{Model selection and estimation in the {G}aussian graphical model}.
\bjournal{Biometrika}
\bvolume{94}
\bpages{19--35}.
\bid{doi={10.1093/biomet/asm018}, issn={0006-3444}, mr={2367824}}
\end{barticle}
%
\bptok{imsref}%
\endbibitem

\bibitem[\protect\citeauthoryear{Zhou and Zhu}{2007}]{ZhouZhu07}
%
\begin{bmisc}[author]
\bauthor{\bsnm{Zhou},~\bfnm{N.}\binits{N.}} \AND
\bauthor{\bsnm{Zhu},~\bfnm{J.}\binits{J.}}
(\byear{2007}).
\bhowpublished{Group variable selection via a hierarchical lasso and
its oracle property.
Technical report,
Dept. Statistics, Univ. Michigan, Ann Arbor, MI}.
\end{bmisc}
%
\bptok{imsref}%
\endbibitem

\bibitem[\protect\citeauthoryear{Zou and Hastie}{2005}]{ZouHastie05}
%
\begin{barticle}[mr]
\bauthor{\bsnm{Zou},~\bfnm{Hui}\binits{H.}} \AND
\bauthor{\bsnm{Hastie},~\bfnm{Trevor}\binits{T.}}
(\byear{2005}).
\btitle{Regularization and variable selection via the elastic net}.
\bjournal{J. R. Stat. Soc. Ser. B Stat. Methodol.}
\bvolume{67}
\bpages{301--320}.
\bid{doi={10.1111/j.1467-9868.2005.00503.x}, issn={1369-7412}, mr={2137327}}
\end{barticle}
%
\bptok{imsref}%
\endbibitem
\end{thebibliography}
\end{document}